\documentclass[twocolumn,secnumarabic,amsmath,amssymb,nobibnotes,aps,prd]{revtex4-2}
\usepackage{graphicx}
\usepackage{slashed}
\usepackage{bbold}
\begin{document}
	\title{Nakanishi integral representation of pseudoscalar fermion-antifermion bound states in the Bethe--Salpeter equation}
	\author{Shaoyang Jia}
	\email[]{syjia@anl.gov}
	\affiliation{Physics Division, Argonne National Laboratory, 9700 S. Cass Avenue, Lemont, IL 60439, USA}
	\date{\today}
	\begin{abstract}
		We propose a method to solve for the pseudoscalar bound state amplitude of a fermion and an antifermion from its Bethe--Salpeter equation (BSE) in the Minkowski space applying spectral representations. With the dressing of the fermion propagators specified by their spectral functions, we first derive the explicit expressions for the Nakanishi spectral functions of the Bethe--Salpeter wave function (BSWF) in terms of these propagators and the Bethe--Salpeter amplitude (BSA) directly from the definition in the momentum space. Applying the rainbow-ladder truncation in the covariant gauge, we then convert the BSE in the momentum space into identities that calculate the spectral functions of the BSA from those of the gauge-boson propagator and of the BSWF. These relations form a closed set of linear functionals for the Nakanishi spectral functions with analytical kernels, laying the foundation for the numerical solution of the BSE in the Minkowski space.
	\end{abstract}
	\maketitle
	\section{Introduction\label{sc:introduction}}
	In terms of Green's function, the structure of $2$-body bound states of a quantum field theory is determined from the Bethe--Salpeter equation (BSE), which requires the propagators of the constituent fermions and their interactions as inputs~\cite{Maris:1999nt}. The fermion propagators are to be solved from their Schwinger-Dyson equation (SDE) with the same interactions~\cite{Windisch:2016iud,Jia:2017niz}. While the vertices of the interactions satisfy their SDEs, which involve higher $n$-point functions than themselves~\cite{Eichmann:2014xya}. For mesons as relativistic bound states of quantum chromodynamics (QCD) interactions of a quark and an antiquark, a practical scheme to truncate these coupled integral equations is the rainbow-ladder, which includes the Maris-Tandy modeling of the quark-gluon interaction in the Landau gauge~\cite{Maris:1999nt}. After adopting a truncation scheme, the resulting integral equations are usually solved in the Euclidean space after the Wick rotation at a given frame of reference~\cite{Jia:2024dfl}. The Euclidean-space versions of these equations are relatively straightforward to solve numerically because singularities of the Green's functions are not expected to present with spacelike momenta~\cite{Windisch:2016iud}. Some of these functions may however become complex-valued while sampling complex momenta~\cite{Jia:2024etj}. 
	
	As integral equations of Green's functions with full kinematics, both the SDE and the BSE are originally formulated in the Minkowski space~\cite{Schwinger:1951ex,Schwinger:1951hq,Dyson:1949ha}. Because the Green's functions are expected to contain singularities for timelike momenta, the Minkowski-space formulation of these equations can be facilitated by the spectral representations. These representations express the Green's functions in terms of their discontinuities, among which are the K\"all\'en--Lehmann spectral representation for the dressed propagator of scalar particles and the Nakanishi integral representation of the Bethe--Salpeter amplitude (BSA)~\cite{nakanishi1971graph,Kusaka:1995za,Kusaka:1997xd,Jia:2023imt}. There also exist the spectral representation of quantum electrodynamics (QED) fermion propagators and the associated truncation for their SDE~\cite{Delbourgo:1977jc,Salam:1963sa,Salam:1964zk,Strathdee:1964zz,Jia:2017niz}. 
	
	Although indirect methods of solving for the Nakanishi spectral functions from the BSE exist, we find it unnecessary to develop a framework based on them~\cite{Carbonell:2010zw,Carbonell:2010tz,Frederico:2013vga,dePaula:2016oct,Karmanov:2021imh}. Our method is instead based on the native formulation of the momentum-space BSE. For scalar bound states of $2$ scalar particles, such a formulation in the Minkowski space with numerical integration kernel functions has been developed in Refs.~\cite{Kusaka:1995za,Kusaka:1997xd}. A subsequent development with analytical kernel functions and the support of dressed propagators is given in Ref.~\cite{Jia:2023imt}, which generates results in agreement with those in Refs.~\cite{Kusaka:1995za,Kusaka:1997xd}. 
	
	While in this article we generalize this analytical method to pseudoscalar bound states of fermions in the rainbow-ladder truncation with the interaction of massless gauge-boson exchange. Specifically we first introduce the Nakanishi integral representations with supporting operations on the spectral functions of both the BSA and the Bethe--Salpeter wave function (BSWF). These supporting operations when properly applied avoid the complication due to mixing differential operators with regular functions in the Nakanishi spectral functions~\cite{Jia:2023imt}. Applying the spectral representation of the fermion propagators, the definition of the BSWF in the momentum space is subsequently converted into an array of integral relations for the spectral functions involved. We also derive an array of linear functionals of Nakanishi spectral functions from the BSE in the rainbow-ladder truncation in the covariant gauge, where the gauge-boson propagator is dressed in the Landau gauge. After knowing the spectral functions of the propagators, the obtained equations form a closed set of linear integral relations for the Nakanishi spectral functions. The solution of the BSE could then be obtained after choosing a normalization condition. Because we work in the Minkowski space, we adopt $g^{\mu\nu} = { \{ 1, -1, -1, \dots , -1\} }$ as the spacetime metric such that the momentum $p^\mu$ is timelike when $p^2 \geq 0$. Although the gauge dependence of the QED fermion propagator is known exactly in the covariant gauge, such a dependence for the solutions of the BSE is beyond the scope of this article~\cite{Jia:2016udu,Jia:2016wyu,Jia:2016aau}.
	
	This article is organized as follows. Section~\ref{sc:introduction} is the introduction. Section~\ref{sc:NIR} introduces the Nakanishi integral representation of pseudoscalar bound state amplitudes with associated elementary operations on the spectral functions. Section~\ref{sc:BSWF} contains the derivation of the functional identities for the spectral functions of the BSWF in terms of those of the fermion propagators and of the BSA. Similar expressions for the spectral functions of the BSA based on its BSE in the rainbow-ladder truncation with dressed gauge-boson propagator are given in Sec.~\ref{sc:BSE}. Section~\ref{sc:summary} contains the conclusion. Supporting equations are given in the Appendix. 
	\section{Nakanishi integral representation\label{sc:NIR}}
	The BSA for pseudoscalar bound states of two fermions decomposes as follows~\cite{Jia:2024dfl}
	\begin{align}
		& \Gamma(k, P) = \gamma_5 [ E(k^2, k\cdot P) + \slashed{P} F(k^2, k\cdot P) \nonumber\\
		& + \slashed{k} G(k^2, k\cdot P) -i \sigma_{kP} H(k^2, k\cdot P) ], \label{eq:def_BSV_ps}
	\end{align}
	where we have introduced ${\sigma_{kP} = k^\mu P^\mu \sigma_{\mu\nu}}$ with the Dirac matrix ${\sigma^{\mu\nu} = i(\gamma^\mu \gamma^\nu - \gamma^\nu \gamma^\mu)/2}$. The momenta of the quark and the antiquark are given by 
	\begin{equation}
		k^\mu_\pm = k^\mu \pm \eta_\pm P^\mu \label{eq:def_k_pm}
	\end{equation}
	with parameters $\eta_\pm$ specifying the partition of the bound state momentum $P^\mu$. Momentum conservation then requires ${\eta_+ + \eta_- = 1}$ as illustrated in Fig.~\ref{fig:bswfps}, which also results in $k^{\mu} = {\eta_- k^{\mu}_+ + \eta_+ k^{\mu}_-}$. After defining the following scalar contractions of the Dirac matrices with momentum variables
	\begin{equation}
		T(k,P) = \left( 1,\, \slashed{P},\, \slashed{k}, -i\sigma_{kP} \right), \label{eq:def_T_j_basis}
	\end{equation}
	the Nakanishi integral representation of the BSA is given by~\cite{nakanishi1971graph,Kusaka:1995za,Kusaka:1997xd,Jia:2023imt}
	\begin{align}
		& \Gamma(k,P) = \gamma_5 \sum_{j=1}^{4} T_j(k,P)\int_{z_{\mathrm{th}}^-}^{z_{\mathrm{th}}^+} dz\int_{g_{\mathrm{th}}(z)}^{+\infty} d\gamma \nonumber\\
		& \times \phi_j(\gamma,z) / (k^2 +zk\cdot P -\gamma +i\varepsilon)^{n_j}. \label{eq:NIR_BSA_ps}
	\end{align}
	Here ${\phi_j(\gamma,z)}$ is the Nakanishi spectral function of the scalar component $j$, with $\gamma$ being the radial variable and $z$ the angular variable. The limits of the angular integration are $z^-_{\mathrm{th}}$ and $z^+_{\mathrm{th}}$, which are usually chosen to be $-1$ and $1$ respectively. Meanwhile the function $g_{\mathrm{th}}(z)$ is the lower threshold of the radial spectral variable at a given value of $z$~\cite{Jia:2023imt}. The index of the denominator $n_j\geq 1$ is unique to each of the scalar components. While the term $i\varepsilon$ stands for the Feynman prescription.
	
	In order to solve for the Nakanishi spectral functions in Eq.~\eqref{eq:NIR_BSA_ps} from the dynamics, we need to convert the BSE in the momentum space into identities of the spectral functions. This can be accomplished by matching the momentum dependence on both sides of the BSE, which requires the translation of momentum-space operations into those on the spectral functions~\cite{Jia:2023imt}. Specifically we expect changing indices of the denominators in Eq.~\eqref{eq:NIR_BSA_ps}, the multiplication of $k\cdot P$, and the multiplication of $k^2$ all to occur for a given scalar component.
	
	For the lowering of the denominator exponent, we need to find a spectral function $\Phi(\gamma,z)$ that is related to the spectral function $\phi(\gamma,z)$ for $n\geq 1$ according to 
	\begin{align}
		& \int dz \int_{g_{\mathrm{th}}(z)}^{+\infty} d\gamma \dfrac{\Phi(\gamma,z)}{(k^2 + zk\cdot P -\gamma)^n} = \int dz \int_{g_{\mathrm{th}}(z)}^{+\infty} d\gamma \nonumber\\
		& \times \phi(\gamma,z) / (k+zk\cdot P - \gamma)^{n+1}, \label{eq:cond_cal_D_minus}
	\end{align}
	where the limits of integration for the angular variable are implicit. Applying the following integration by parts
	\begin{align}
		& \int_{g_{\mathrm{th}}(z)}^{+\infty} d\gamma\, \dfrac{1}{(k^2+zk\cdot P-\gamma)^n} \dfrac{\partial}{\partial \gamma} \phi(\gamma,z) \nonumber\\
		& = \lim\limits_{\gamma\rightarrow +\infty} \dfrac{\phi(\gamma,z)}{(k^2 + zk\cdot P - \gamma)^n} - \dfrac{\phi(g_{\mathrm{th}}(z), z)}{[ k^2 + zk\cdot P -g_{\mathrm{th}}(z) ]^n} \nonumber\\
		& - n \int_{g_{\mathrm{th}}(z)}^{+\infty} d\gamma\, \dfrac{\phi(\gamma,z)}{(k^2 + zk\cdot P - \gamma)^{n+1}},
	\end{align}
	we obtain
	\begin{subequations}\label{eq:def_cal_D_minus}
		\begin{align}
			& \Phi(\gamma,z) = \big\{ [ \delta(\gamma-\infty)-\delta(\gamma-g_{\mathrm{th}}(z)) ] \phi(\gamma,z) \nonumber\\
			& - \partial \phi(\gamma,z) / \partial \gamma \big\} /n .\label{eq:ori_cal_D_minus}
		\end{align}
		Let us then define a linear functional $\mathcal{D}_{-}$ of $\phi(\gamma,z)$ according to
		\begin{align}
			& \mathcal{D}_{-}[\phi](\gamma,z) = \dfrac{1}{n} \left[ \delta( \gamma - \infty) - \delta( \gamma - g_{\mathrm{th}}(z)) - \dfrac{\partial}{\partial \gamma } \right] \nonumber\\
			& \times \phi(\gamma,z),
		\end{align}
	\end{subequations}
	with $n+1$ being the exponent in the denominator of the Nakanishi integral representation using $\phi(\gamma,z)$. The function $\Phi(\gamma,z)$ given by Eq.~\eqref{eq:ori_cal_D_minus} ensures Eq.~\eqref{eq:cond_cal_D_minus} in the momentum space. We have adopted the notation that the function arguments of a functional are enclosed in square brackets, while the variables for the resulting function as a return of the functional are specified in round brackets immediately following the square ones. We also adopt the convention that multiplications of functionals indicate their compound in the order of their multiplications, which is analogous to matrix multiplications for finite-dimensional linear operators. The unit functional is therefore denoted by ``$\mathbb{1}$''. While ``$\mathbb{0}$'' stands for the functional that returns zero.
	
	In contrast to Eq.~\eqref{eq:cond_cal_D_minus}, for the raising of the denominator exponent we would like to find another spectral function $\Phi(\gamma,z)$ that is related to $\phi(\gamma,z)$ when $n\geq 1$ via
	\begin{align}
		& \int dz \int_{g_{\mathrm{th}}(z)}^{+\infty} d\gamma \dfrac{\Phi(\gamma,z)}{(k^2 + zk\cdot P - \gamma)^{n+1}} = \int dz \int_{g_{\mathrm{th}}(z)}^{+\infty} d\gamma \nonumber\\
		& \times \phi(\gamma,z) / (k^2 + zk\cdot P - \gamma)^n. \label{eq:cond_cal_D_plus}
	\end{align}
	Here limits of the angular integration are again implicit. The application of integration by parts then gives
	\begin{align}
		& \int_{g_{\mathrm{th}}(z)}^{+\infty} d\gamma \dfrac{ 1 }{ (k^2+zk\cdot P-\gamma)^{n+1} } \int_{g_{\mathrm{th}}(z)}^{\gamma} d\beta\, \phi(\beta, z) \nonumber\\
		& = \dfrac{1}{n}\Big\{ \lim\limits_{\gamma\rightarrow +\infty} \dfrac{1}{(k^2+zk\cdot P - \gamma)^{n} } \int_{g_{\mathrm{th}}(z)}^{\gamma} d\beta\,\phi(\beta,z) \nonumber\\
		& - \int_{g_{\mathrm{th}}(z)}^{+\infty} d\gamma \, \phi(\gamma,z) / (k^2+zk\cdot P - \gamma)^n \Big\},
	\end{align}
	which indicates
	\begin{subequations}\label{eq:def_cal_D_plus}
	\begin{equation}
		\Phi(\gamma,z) = - n \int_{g_{\mathrm{th}}(z)}^{\gamma}d\beta\, \phi(\beta,z), \label{eq:ori_cal_D_plus}
	\end{equation}
	taking the assumption that
	\begin{equation*}
		\lim\limits_{\gamma\rightarrow +\infty} \dfrac{1}{(k^2+zk\cdot P - \gamma)^{n}} \int_{g_{\mathrm{th}}(z)}^{\gamma} d\beta\,\phi(\beta,z) = 0.
	\end{equation*}
	This condition restricts the asymptotic behavior of the spectral function $\phi(\gamma,z)$ in the radial variable. Specifically when $\lim\limits_{\gamma\rightarrow +\infty} \phi(\gamma,z) \sim \gamma ^{ - \alpha }$, it requires that ${\alpha + n > 1}$. When such a condition can be satisfied, we define the corresponding linear functional $\mathcal{D}_{+}$ as
	\begin{equation}
		\mathcal{D}_{+}[\phi](\gamma,z) = -n \int_{g_{\mathrm{th}}(z)}^{\gamma}d\beta \, \phi(\beta,z),
	\end{equation}
	\end{subequations}
	where $n$ is the exponent in the denominator of the Nakanishi integral representation using $\phi(\gamma,z)$. Equation~\eqref{eq:ori_cal_D_plus} then gives the spectral function $\Phi(\gamma,z)$ that ensures Eq.~\eqref{eq:cond_cal_D_plus} in the momentum space. Notice that the operators $\mathcal{D}_-$ and $\mathcal{D}_+$ are mutually inverse. Although there is no extra requirement to ensure $\mathcal{D}_+\mathcal{D}_- = \mathbb{1}$, conditions to maintain $\mathcal{D}_-\mathcal{D}_+ = \mathbb{1}$ are either ${\int_{g_{\mathrm{th}}(z)}^{+\infty}d\beta \, \phi(\beta,z)} = 0$ or $\lim\limits_{\gamma \rightarrow +\infty } { \int_{ z^-_{\mathrm{th}} }^{ z^+_{\mathrm{th}} } dz (k^2 + zk\cdot P -\gamma )^{-n}} = 0$, the later of which is satisfied when $n\geq1$. 
	
	We may also encounter a scalar component multiplied by $k\cdot P$ in the momentum space after the Nakanishi integral representation, which corresponds to finding a spectral function $\Phi(\gamma,z)$ related to $\phi(\gamma,z)$ for $n\geq 1$ by 
	\begin{align}
		& \int d\gamma \int_{z_{\mathrm{th}}^-(\gamma)}^{z_{\mathrm{th}}^+(\gamma)} dz \dfrac{\Phi(\gamma,z)}{(k^2 + zk\cdot P -\gamma)^n} = k\cdot P \int d\gamma \nonumber\\
		& \times \int_{z_{\mathrm{th}}^-(\gamma)}^{z_{\mathrm{th}}^+(\gamma)} dz \dfrac{\phi(\gamma,z)}{(k^2 + zk\cdot P - \gamma)^{n+1}}. \label{eq:cond_cal_P_minus}
	\end{align}
	Here the integration limits for $\gamma$ are implicit. Notice that we have exchanged the order of integrations for $\gamma$ and $z$, which results in the thresholds $z^\pm_{\mathrm{th}}$ depending on the spectral variable $\gamma$. Another application of integration by parts gives 
	\begin{align}
		& \int_{z_{\mathrm{th}}^-(\gamma)}^{z_{\mathrm{th}}^+(\gamma)}dz \dfrac{k\cdot P\, \phi(\gamma,z)}{(k^2 + zk\cdot P -\gamma)^{n+1}} = \dfrac{1}{n} \bigg\{ \nonumber\\
		& - \dfrac{ \phi(\gamma,z_{\mathrm{th}}^+(\gamma)) }{[k^2+z_{\mathrm{th}}^+(\gamma)k\cdot P - \gamma]^n} + \dfrac{ \phi(\gamma,z_{\mathrm{th}}^-(\gamma)) }{[k^2+z_{\mathrm{th}}^-(\gamma)k\cdot P - \gamma]^n} \nonumber\\
		& + \int_{z_{\mathrm{the}}^-(\gamma)}^{z_{\mathrm{th}}^+(\gamma)} dz\, \dfrac{1}{ (k^2 + zk\cdot P - \gamma )^n } \dfrac{ \partial }{\partial z}\phi(\gamma,z) \bigg\}.
	\end{align}
	Consequently we obtain
	\begin{subequations}
		\begin{align}
			& \Phi(\gamma,z) = \big\{ [ \delta( z-z_{\mathrm{th}}^-(\gamma) ) - \delta( z-z_{\mathrm{th}}^+(\gamma) ) ] \, \phi(\gamma,z) \nonumber\\
			& + \partial \phi(\gamma,z) / \partial z \big\} /n  \label{eq:ori_cal_P_minus}
		\end{align}
		with $n+1$ being the exponent in the denominator of the Nakanishi integral representation using $\phi(\gamma,z)$ as in Eq.~\eqref{eq:cond_cal_P_minus}. Equation~\eqref{eq:ori_cal_P_minus} yields the spectral function $\Phi(\gamma,z)$ that satisfies Eq.~\eqref{eq:cond_cal_P_minus}, in association with which let us define a linear functional $\mathcal{P}_{-}$ of $\phi(\gamma,z)$ as
		\begin{align}
			& \mathcal{P}_{-}[\phi](\gamma,z) = \dfrac{1}{n}\left[ \delta(z-z_{\mathrm{th}}^-(\gamma)) - \delta(z-z_{\mathrm{th}}^+(\gamma)) + \dfrac{\partial}{\partial z} \right] \nonumber\\
			& \times \phi(\gamma,z). \label{eq:def_cal_P_minus}
		\end{align}
	\end{subequations}
	Dirac $\delta$-functions in Eq.~\eqref{eq:def_cal_P_minus} can be converted into those of the radial variable $\gamma$ located at the radial threshold $\gamma_{\mathrm{th}}(s)$ and the angular parts located at $z^{\pm}_{\mathrm{th}}$ as the global extrema of the angular variable $z$. Explicitly when $z^{\pm}_{\mathrm{th}}(\gamma)$ are distributed on separate sides of $z_0$ for any given $\gamma$ such that ${ z^{-}_{\mathrm{th}}(\gamma) \leq z_0 \leq z^{+}_{\mathrm{th}}(\gamma) }$, we have
	\begin{align}
		& \delta(z-z_{\mathrm{th}}^-(\gamma)) - \delta(z-z_{\mathrm{th}}^+(\gamma)) = \delta(z-z_{\mathrm{th}}^-) - \delta(z-z_{\mathrm{th}}^+) \nonumber\\
		& - \mathrm{sign}( z - z_0 ) \left\vert \dfrac{ d\gamma_{\mathrm{th}}(z)}{dz} \right\vert \delta( \gamma - \gamma_{\mathrm{th}}(z) ).\label{eq:delta_z_pm_rdd}
	\end{align}
	Furthermore when the sign of the second-order derivative $d^2 \gamma_{\mathrm{th}}(z) / dz^2$ is known, the coefficient of the radial $\delta$-function in Eq.~\eqref{eq:delta_z_pm_rdd} can be reduced based on
	\begin{subequations}
		\begin{align}
			& g(z) = \mathrm{sign}( z - z_0 ) \left\vert \dfrac{ d\gamma_{\mathrm{th}}(z)}{dz} \right\vert = \mathrm{sign}\left( \dfrac{ d^2\gamma_{\mathrm{th}}(z)}{dz^2} \right) \dfrac{d\gamma_{\mathrm{th}}(z)}{dz} \nonumber\\
			& =
			\begin{cases}
				\dfrac{d\gamma_{\mathrm{th}}(z)}{dz} \quad (\text{for convex}~\gamma_{\mathrm{th}}(z) ) \\[2mm]
				- \dfrac{d\gamma_{\mathrm{th}}(z)}{dz} \quad (\text{for concave}~\gamma_{\mathrm{th}}(z) )
			\end{cases}.
		\end{align}
		In this case the linear functional $\mathcal{P}_{-}$ defined in Eq.~\eqref{eq:def_cal_P_minus} becomes
		\begin{align}
			& \mathcal{P}_{-}[\phi](\gamma,z) = \dfrac{1}{n} \big[ g(z)\, \delta( \gamma - \gamma_{\mathrm{th}}(z)) + \delta(z-z_{\mathrm{th}}^-) \nonumber\\
			& - \delta(z-z_{\mathrm{th}}^+) + \partial / \partial z \big] \phi(\gamma,z). \label{eq:cal_P_minus_rdd}
		\end{align}
	\end{subequations}
	
	Similar to the case of multiplication by $k\cdot P$, we would also like to find a spectral function $\Phi(\gamma,z)$ that is related to the function $\phi(\gamma,z)$ for $n\geq 1$ by 
	\begin{align}
		& \int_{z_{\mathrm{th}}^-}^{z_{\mathrm{th}}^+}dz \int_{g_{\mathrm{th}}(z)}^{+\infty} d\gamma\, \dfrac{\Phi(\gamma,z)}{(k^2+zk\cdot P -\gamma)^n} = k^2 \int_{z_{\mathrm{th}}^-}^{z_{\mathrm{th}}^+}dz \nonumber\\
		& \times \int_{g_{\mathrm{th}}(z)}^{+\infty} d\gamma\, \dfrac{\phi(\gamma,z)}{(k^2+zk\cdot P -\gamma)^{n+1}}. \label{eq:cond_cal_K_minus}
	\end{align}
	The identity
	\begin{equation}
		k^2/(k^2+zk\cdot P - \gamma) = 1 + {( \gamma -zk\cdot P )}/{(k^2+zk\cdot P - \gamma )} \label{eq:dec_k2_multi}
	\end{equation}
	indicates the following result
	\begin{align}
		& \int_{z_{\mathrm{th}}^-}^{z_{\mathrm{th}}^+}dz \int_{g_{\mathrm{th}}(z)}^{+\infty} d\gamma \dfrac{k^2 \phi(\gamma,z)}{ (k^2 + zk\cdot P -\gamma)^{n+1}} = \int_{z_{\mathrm{th}}^-}^{z_{\mathrm{th}}^+}dz \nonumber\\
		& \int_{g_{\mathrm{th}}(z)}^{+\infty} d\gamma \dfrac{1}{(k^2 + zk\cdot P -\gamma)^n} \Big\{ \phi(\gamma,z) + \dfrac{1}{n} \big\{ [ \delta(z-z_{\mathrm{th}}^+(\gamma)) \nonumber\\
		& - \delta(z-z_{\mathrm{th}}^-(\gamma)) ] z\, \phi(\gamma,z) - \frac{\partial}{\partial z} z\, \phi(\gamma,z) + [ \delta(\gamma-\infty) \nonumber\\
		& - \delta( \gamma - g_{\mathrm{th}}(z) ) ] \gamma\, \phi(\gamma,z) - \frac{\partial}{\partial \gamma } \gamma\, \phi(\gamma,z) \big\} \Big\} . 
	\end{align}
	Consequently we obtain 
	\begin{subequations}\label{eq:def_cal_K_minus}
	\begin{align}
		& \Phi(\gamma,z) = \phi(\gamma,z) + \dfrac{1}{n} \Big\{ [\delta(\gamma-\infty) - \delta(\gamma-g_{\mathrm{th}}(z))] \gamma \nonumber\\
		& \times \phi(\gamma,z) - \dfrac{\partial}{\partial \gamma} \gamma\, \phi(\gamma,z) + [ \delta(z-z_{\mathrm{th}}^+(\gamma)) - \delta(z-z_{\mathrm{th}}^-(\gamma)) ] \nonumber\\
		& \times z\, \phi(\gamma,z) - \dfrac{\partial}{\partial z} z\, \phi(\gamma,z) \Big\} \label{eq:multi_k2_form}, 
	\end{align}
	where ${n+1}$ is the exponent of the denominator in the Nakanishi integral representation using $\phi(\gamma,z)$. Equation~\eqref{eq:multi_k2_form} defines a linear functional $\mathcal{K}_-$ that gives the function $\Phi(\gamma,z)$ in terms of $\phi(\gamma,z)$ formally as
	\begin{equation}
		\Phi(\gamma,z) = \mathcal{K}_-[\phi](\gamma,z)
	\end{equation}
	in order to satisfy Eq.~\eqref{eq:cond_cal_K_minus} in the momentum space. Applying previously defined operators, we further obtain
	\begin{equation} 
		\mathcal{K}_{-}[\phi](\gamma,z) = \mathbb{1}_{-}[\phi](\gamma,z) - \mathcal{P}_{-}[z\, \phi](\gamma,z) + \mathcal{D}_{-}[\gamma\,\phi](\gamma,z)
	\end{equation}
	\end{subequations}
	as indicated by Eq.~\eqref{eq:dec_k2_multi}. Here $\mathbb{1}_{-}$ is the operator that only lowers the denominator index by unity. One could further demonstrate that the commutators of operators $\mathcal{D}_{-}$, $\mathcal{D}_{+}$, $\mathcal{P}_{-}$, and $\mathcal{K}_{-}$ defined in Eqs.~\eqref{eq:def_cal_D_minus}, \eqref{eq:def_cal_D_plus}, \eqref{eq:def_cal_P_minus}, and \eqref{eq:def_cal_K_minus} all vanish. These operators are the building blocks of functional identities for the Nakanishi spectral functions to be developed.
	\section{Bethe--Salpeter wave function\label{sc:BSWF}}
	\subsection{Scalar decomposition}
	\begin{figure}
		\centering
		\includegraphics[width=\linewidth]{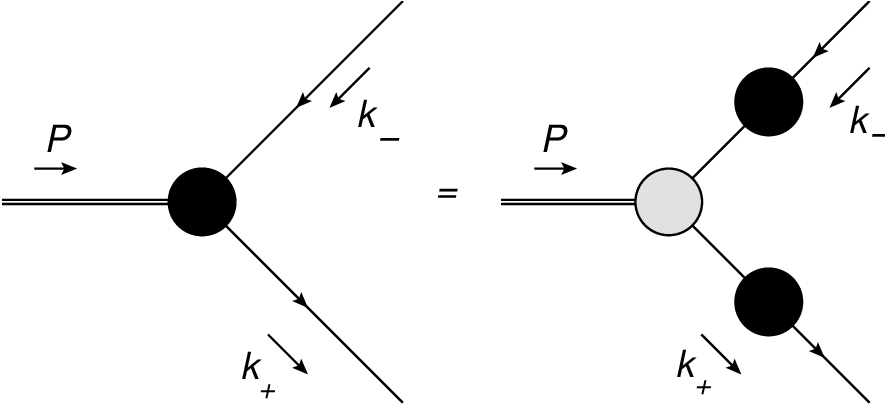}
		\caption{Definition of the BSWF as products of the BSA and fermion propagators. The black blob and the gray blob both with $3$ external legs are respectively the BSWF and the BSA. The remaining black blobs are the dressed fermion propagators. The momenta of the bound state and of the quarks are respectively $P^\mu$ and $k^\mu_\pm$.}
		\label{fig:bswfps}
	\end{figure}
	The BSWF in the momentum space is defined in relation to the BSA by 
	\begin{equation}
		\chi(k, P) = S^+_{\mathrm{F}}(k_+) \Gamma(k,P) S^-_{\mathrm{F}}(k_-), \label{eq:def_BSE_wv_ps}
	\end{equation}
	with $S^\pm_{\mathrm{F}}(k_\pm)$ being the propagators of the valence quarks~\cite{Maris:1999nt}. This definition is diagrammatically represented in Fig.~\ref{fig:bswfps}. Similar to the BSA in Eq.~\eqref{eq:def_BSV_ps}, the BSWF can be decomposed into 4 scalar components according to
	\begin{align}
		& \chi(k,P) = \gamma_5 [ E_{\chi}(k^2, k\cdot P) + \slashed{P} F_{\chi}(k^2, k\cdot P) \nonumber\\
		& + \slashed{k} G_{\chi}(k^2, k\cdot P) - i\sigma_{kP} H_{\chi}(k^2, k\cdot P) ]. \label{eq:scalar_decomposition_BSWF_ps}
	\end{align}
	In order to access the the BSWF in Eq.~\eqref{eq:def_BSE_wv_ps} for timelike momenta, let us introduce the spectral representation of the quark propagators through
	\begin{equation}
		S_{\mathrm{F}}(k) = \int_{\vert W \vert \geq W_{\mathrm{th}}}^{+\infty} dW \, \dfrac{\rho(W)}{\slashed{k}-W +i\varepsilon \, \mathrm{sign}(W)}, \label{eq:SR_S_F}
	\end{equation}
	the scalar decomposition of which is explained in the Appendix~\cite{Jia:2016wyu,Jia:2017niz}. Meanwhile the Nakanishi integral representation of the BSWF is given by 
	\begin{align}
		& \chi(k,P) = \gamma_5\sum_{j=1}^{4} T_j(k,P) \int_{z_{\mathrm{th}}^-}^{z_{\mathrm{th}}^+}dz \int_{\Gamma_{\mathrm{th}}(z)}^{+\infty} d\gamma \nonumber\\
		& \times \Phi_j(\gamma,z) / (k^2+zk\cdot P-\gamma +i\varepsilon)^{m_j}, \label{eq:NIR_BSWF_ps}
	\end{align}
	where $\Phi_j(\gamma,z)$ is the Nakanishi spectral function for the $j$-th component with an associated integer $m_j$. We expect the indices in the denominators to be related to those of the BSA by ${m_j= n_j+1}$ based on the momentum dimensions in Eq.~\eqref{eq:def_BSE_wv_ps}. The lower limit of integration $\Gamma_{\mathrm{th}}(z)$ for the radial variable remains to be determined. Substituting Eqs.~\eqref{eq:NIR_BSA_ps} and~\eqref{eq:SR_S_F} into the definition of the BSWF in Eq.~\eqref{eq:def_BSE_wv_ps} then gives
	\begin{align}
		& \chi(k,P) = \sum_{j=1}^{4} \int_{\vert V \vert \geq V_{\mathrm{th}}}^{+\infty} dV \int_{\vert W \vert \geq W_{\mathrm{th}}}^{+\infty} dW \int_{z^-_{\mathrm{th}}}^{z^+_{\mathrm{th}}} dz  \int_{g_{\mathrm{th}(z)}}^{+\infty} d\gamma \nonumber\\
		& \times \dfrac{ (\slashed{k}_+ + V) \gamma_5 T_j(k,P) (\slashed{k}_- + W) \rho_+(V) \rho_-(W) \phi_j(\gamma,z) }{ (k_+^2-V^2) (k_-^2-W^2) (k^2+zk\cdot P-\gamma +i\varepsilon)^{n_j} }, \label{eq:chi_in_rho_phi}
	\end{align}
	where $\rho_+(V)$ and $\rho_-(W)$ are respectively the spectral functions of the quark and the antiquark propagators with corresponding thresholds of $V_{\mathrm{th}}$ and $W_{\mathrm{th}}$. Equations~\eqref{eq:NIR_BSWF_ps} and~\eqref{eq:chi_in_rho_phi} indicate that $\Phi_i(\gamma,z)$ can be written as a linear functional in both $\rho_\pm(W)$ and $\phi_j(\gamma,z)$ in order to satisfy the momentum-space identity in Eq.~\eqref{eq:def_BSE_wv_ps}. Because all scalar components need to match, we expand the momentum factors in the numerator of Eq.~\eqref{eq:chi_in_rho_phi} in terms of the Dirac bases defined in Eq.~\eqref{eq:def_T_j_basis}. The corresponding result for each component is obtained with the assistance of the following function:
	\begin{equation}
		N(\gamma_5 T_j(k,P)) = (\slashed{k}_++V)\, \gamma_5 T_j(k,P)\, (\slashed{k}_-+W),
	\end{equation}
	with $k^\mu_\pm$ defined in Eq.~\eqref{eq:def_k_pm}. Specifically with this definition of $N(\gamma_5 T_j)$ we obtain
	\begin{subequations}\label{eq:momentum_merge_components_ps}
		\begin{align}
			& N(\gamma_5) = \gamma_5 [ (\eta_- - \eta_+) k\cdot P - k^2 + \eta_+ \eta_- P^2 + VW \nonumber\\
			& - (\eta_-V+\eta_+W)\slashed{P} + (V-W)\slashed{k} - i\sigma_{kP} ], \\
			& N(\gamma_5\slashed{P}) = \gamma_5 \big\{ (V-W) k\cdot P - (\eta_-V+\eta_+W) P^2 \nonumber\\
			& + ( k^2 + \eta_+ \eta_- P^2 +VW )\slashed{P} - [ 2k\cdot P + (\eta_+ - \eta_-)P^2 ] \slashed{k} \nonumber \\ 
			& + (V+W)i\,\sigma_{kP} \big\}, \\
			& N(\gamma_5\slashed{k}) = \gamma_5 \big\{ (V-W) k^2 - ( \eta_-V + \eta_+W ) k\cdot P \nonumber\\
			& + [ 2\eta_+ \eta_- k\cdot P - (\eta_+ - \eta_-) k^2 ] \slashed{P} + [ VW \nonumber\\
			& - (k^2 + \eta_+\eta_-P^2) ] \slashed{k} + ( \eta_- V - \eta_+ W ) i\, \sigma_{kP} ) \big\},
		\end{align}
		and
		\begin{align}
			& N(\gamma_5(-i)\sigma_{kP}) = \gamma_5 \big\{ k^2P^2 - (k\cdot P)^2 + [ - ( V + W ) k^2 \nonumber\\
			& + ( \eta_-V - \eta_+W ) k\cdot P ] \slashed{P} + [ (V+W) k\cdot P - ( \eta_- V \nonumber\\
			& - \eta_+ W ) P^2 ] \slashed{k} + [ k^2 + (\eta_+ - \eta_-) k\cdot P - \eta_+ \eta_- P^2 \nonumber\\
			& + VW ] (-i)\sigma_{kP} \big\}.
		\end{align}
	\end{subequations}
	\subsection{Linear operators in the scalar BSWF}
	Before matching the momentum dependence for each of the scalar components indicated by Eq.~\eqref{eq:momentum_merge_components_ps}, let us first consider the contribution to a single component of the BSWF from a given component $\psi(k^2,k\cdot P)$ of the BSA. As discussed in Ref.~\cite{Jia:2023imt}, in the momentum space we have 
	\begin{equation}
	\chi(k^2, k\cdot P) = D_{+}(k^2_+) \psi(k^2, k\cdot P) D_{-}(k^2_-)
	\end{equation}
	with $\chi(k^2, k\cdot P)$ standing for a scalar component of the BSWF in Eq.~\eqref{eq:scalar_decomposition_BSWF_ps}. The spectral representations of the scalar propagators $D_\pm(k_\pm^2)$ are given by 
	\begin{equation}
	D_\pm(k_\pm^2) = \int_{s_{\mathrm{th}}}^{+\infty} ds\, \dfrac{\rho_\pm(s)}{k_\pm^2-s+i\varepsilon}.
	\end{equation}
	Meanwhile the Nakanishi integral representation of $\psi(k^2,k\cdot P)$ as a scalar component of the BSA is given by
	\begin{equation}
	\psi(k^2,k\cdot P) = \int_{z_{\mathrm{th}}^-}^{z_{\mathrm{th}}^+}dz \int_{g_{\mathrm{th}}(z)}^{+\infty} d\gamma \dfrac{\phi(\gamma,z)}{(k^2 + zk\cdot P -\gamma)^n}. \label{eq:NIR_BS_vertex_scalar}
	\end{equation}
	While the corresponding representation of the BSWF is
	\begin{equation}
	\chi(k^2,k\cdot P) = \int_{z_{\mathrm{th}}^-}^{z_{\mathrm{th}}^+}dz \int_{\Gamma_{\mathrm{th}}(z)}^{+\infty} d\gamma \dfrac{\Phi(\gamma,z)}{(k^2 + zk\cdot P -\gamma)^{n+2}}, \label{eq:NIR_BSWF_scalar}
	\end{equation}
	where the exponent in the denominator is chosen to match that in Eq.~\eqref{eq:NIR_BS_vertex_scalar}. Based on the explicit calculation that produces Eq.~(14) of Ref.~\cite{Jia:2023imt}, we have
	\begin{subequations}\label{eq:merge_NSR}
	\begin{align}
	& \Phi(\gamma,z) = n(n+1) \int_{r_{\mathrm{th}}}^{+\infty} dr \int_{s_{\mathrm{th}}}^{+\infty} ds \, \rho_+(s) \rho_-(r) \nonumber\\
	& \times [ J(s,r,z,\gamma) ]^{n-1} \bigg\{ \theta( J(s,r,z,\gamma) ) \bigg[ \int_{z_{\mathrm{th}}^-}^{z}dz' \nonumber\\
	& \times \int_{0}^{\gamma_{\mathrm{th}}(-z,\gamma,s,-z')}d\gamma' + \int_{z}^{z_{\mathrm{th}}^+}dz' \int_{0}^{\gamma_{\mathrm{th}}(z,\gamma,r,z')} d\gamma' \bigg] \nonumber\\
	& - \theta( -J(s,r,z,\gamma) ) \bigg[ \int_{z_{\mathrm{th}}^-}^{z}dz' \int_{\gamma_{\mathrm{th}}(-z,\gamma,s,-z')}^{+\infty} d\gamma' \nonumber\\
	& + \int_{z}^{z_{\mathrm{th}}^+}dz' \int_{\gamma_{\mathrm{th}}(z,\gamma,r,z')}^{+\infty}d\gamma' \bigg] \bigg\} \dfrac{\phi(\gamma',z')}{[ J(s,r,z',\gamma') ]^n}, 
	\end{align}
	where we have defined the signed Jacobian
	\begin{equation}
		J(s,r,z,\gamma) = s + r + ( s - r ) z - 2 ( \gamma + P^2/4 )
	\end{equation}
	and the radial threshold
	\begin{equation}
		\gamma_{\mathrm{th}}(z,\gamma,u,z') = \dfrac{1+z'}{1+z}\gamma -\dfrac{z'-z}{1+z} \left( u - \dfrac{P^2}{4} \right).
	\end{equation}
	Meanwhile $s_{\mathrm{th}}$ and $r_{\mathrm{th}}$ are respectively the thresholds for the spectral functions $\rho_+(s)$ and $\rho_-(r)$. Equation~\eqref{eq:merge_NSR} gives the function $\Phi(\gamma,z)$ as a linear functional $\mathcal{M}_{+2}$ of $\rho_+(r)$, $\rho_-(s)$, and $\phi(\gamma,z)$ formally by
	\begin{equation}
	\Phi(\gamma,z) = \mathcal{M}_{+2}[\rho_+, \rho_-, \phi ](\gamma,z). \label{eq:def_functional_M}
	\end{equation}
	\end{subequations}
	The threshold $\Gamma_{\mathrm{th}}(z)$ in Eq.~\eqref{eq:NIR_BSWF_scalar} is implied by the step functions $\theta(x)$ in Eq.~\eqref{eq:merge_NSR}.
	\subsection{Matrix of linear functionals}
	For notational conveniences, let us define a space holder ``$\sqcup$'' that occurs in the arguments of a functional. The argument represented by the space holder is automatically filled by the function immediately following the functional, or explicitly according to
	\begin{equation}
		\mathcal{F}[\sqcup] \phi(\gamma,z) = \mathcal{F}[\phi](\gamma,z) \label{eq:def_spaceholders}
	\end{equation}
	for a functional $\mathcal{F}$. In the case where the arguments of the function are multiplying the space holder, the corresponding multiplication is first carried with respect to the function argument of the functional. As an example we have ${\mathcal{F}[ z^n\sqcup]\phi(\gamma,z) = \mathcal{F}[\phi'](\gamma,z)}$ with ${\phi'(\gamma,z) = z^n\phi(\gamma,z)}$.
	
	Having reproduced the relation that merges spectral functions of the propagators and of BSA into that of the BSWF for a given scalar component, Eqs.~\eqref{eq:NIR_BSWF_ps} and~\eqref{eq:chi_in_rho_phi} indicate that the Nakanishi spectral functions for a pseudoscalar bound state of fermions in Eq.~\eqref{eq:NIR_BSWF_ps} are given by
	\begin{subequations}\label{eq:def_mbb_M}
	\begin{equation}
		\Phi_j(\gamma,z) = \sum_{k=1}^{4}\mathbb{M}_{jk}[\phi_k](\gamma,z),
	\end{equation}
	or in the matrix form as
	\begin{equation*}
		\begin{pmatrix}
			\Phi_1(\gamma,z)\\
			\Phi_2(\gamma,z)\\
			\Phi_3(\gamma,z)\\
			\Phi_4(\gamma,z)
		\end{pmatrix} = 
		\begin{pmatrix}
			\mathbb{M}_{11}[] & \mathbb{M}_{12}[] & \mathbb{M}_{13}[] & \mathbb{M}_{14}[] \\
			\mathbb{M}_{21}[] & \mathbb{M}_{22}[] & \mathbb{M}_{23}[] & \mathbb{M}_{24}[] \\
			\mathbb{M}_{31}[] & \mathbb{M}_{32}[] & \mathbb{M}_{33}[] & \mathbb{M}_{34}[] \\
			\mathbb{M}_{41}[] & \mathbb{M}_{42}[] & \mathbb{M}_{43}[] & \mathbb{M}_{44}[]
		\end{pmatrix}
		\begin{pmatrix}
			\phi_1(\gamma,z) \\
			\phi_2(\gamma,z) \\
			\phi_3(\gamma,z) \\
			\phi_4(\gamma,z)
		\end{pmatrix},
	\end{equation*}
	where $\mathbb{M}_{jk}[]$ is an array of linear operators with the space holder implied in the square brackets. Specifically based on Eq.~\eqref{eq:momentum_merge_components_ps} we obtain
	\begin{align}
		& \mathbb{M}_{11}[\sqcup] = \mathcal{D}_{+}^{m_1-n_1-1} \{ \{ (\eta_- - \eta_+) \mathcal{P}_- - \mathcal{K}_{-} + \eta_+ \eta_- P^2\mathcal{D}_- \} \nonumber\\
		& \times \mathcal{M}_{+2}[1,1,\sqcup] + \mathcal{D}_- \mathcal{M}_{+2}[2,2,\sqcup] \}.
	\end{align}
	Here we have defined ${ \mathcal{M}_{+2}[j,k,\sqcup] = \mathcal{M}_{+2}[\rho_{+j},\rho_{-k},\sqcup] }$ for $j$ and $k\in\{1,2\}$. Recall that the multiplication of operators indicates their compound. Factors of $\mathcal{D}_{+}$ are introduced to compensate for possible mismatches in the indices of denominators. To avoid confusion with the notation of function arguments for functionals, square brackets are not used when combining terms of a given factor. We also obtain for other components that
	\begin{align}
		& \mathbb{M}_{21}[\sqcup] = - \mathcal{D}_{+}^{m_2-n_1-2} ( \eta_- \mathcal{M}_{+2}[2,1,\sqcup] + \eta_+ \mathcal{M}_{+2}[1,2,\sqcup] ), \\
		& \mathbb{M}_{31}[\sqcup] = \mathcal{D}_{+}^{m_3-n_1-2} ( \mathcal{M}_{+2}[2,1,\sqcup] - \mathcal{M}_{+2}[1,2,\sqcup] ), \\
		& \mathbb{M}_{41}[\sqcup] = \mathcal{D}_+^{m_4-n_1-2} \mathcal{M}_{+2}[1,1,\sqcup], \\
		& \mathbb{M}_{12}[\sqcup] = \mathcal{D}_{+}^{m_1-n_2-1} \{ \mathcal{P}_- \left( \mathcal{M}_{+2}[2,1,\sqcup] - \mathcal{M}_{+2}[1,2,\sqcup] \right) \nonumber\\
		& - P^2 \mathcal{D}_- ( \eta_- \mathcal{M}_{+2}[2,1,\sqcup] + \eta_+ \mathcal{M}_{+2}[1,2,\sqcup] ) \}, \\
		& \mathbb{M}_{22}[\sqcup] = \mathcal{D}_+^{m_2 - n_2 - 1} \{ ( \mathcal{K}_- + \eta_+ \eta_- P^2 \mathcal{D}_- ) \mathcal{M}_{+2}[1,1,\sqcup] \nonumber\\
		&  + \mathcal{D}_- \mathcal{M}_{+2}[2,2,\sqcup] \}, \\
		& \mathbb{M}_{32}[\sqcup] = - \mathcal{D}_+^{m_3-n_2-1} \{ 2 \mathcal{P}_- + (\eta_+ - \eta_-)P^2\mathcal{D}_- \} \nonumber\\
		& \times \mathcal{M}_{+2}[1,1,\sqcup], \\
		& \mathbb{M}_{42}[\sqcup] = - \mathcal{D}_+^{m_4-n_2-2} ( \mathcal{M}_{+2}[2,1,\sqcup] + \mathcal{M}_{+2}[1,2,\sqcup] ), \\
		& \mathbb{M}_{13}[\sqcup] = \mathcal{D}_+^{m_1-n_3-1} \{ \mathcal{K}_- ( \mathcal{M}_{+2}[2,1,\sqcup] - \mathcal{M}_{+2}[1,2,\sqcup] ) \nonumber\\
		& - \mathcal{P}_- ( \eta_- \mathcal{M}_{+2}[2,1,\sqcup] + \eta_+ \mathcal{M}_{+2}[1,2,\sqcup] ) \}, \\
		& \mathbb{M}_{23}[\sqcup] = \mathcal{D}_+^{m_2-n_3-1} \{ 2\eta_+ \eta_- \mathcal{P}_- - (\eta_+ -\eta_-) \mathcal{K}_- \} \nonumber\\
		& \times \mathcal{M}_{+2}[1,1,\sqcup], \\
		& \mathbb{M}_{33}[\sqcup] = \mathcal{D}_+^{m_3 - n_3 - 1} \{ - ( \mathcal{K}_- + \eta_+\eta_-P^2 \mathcal{D}_- ) \mathcal{M}_{+2}[1,1,\sqcup] \nonumber\\
		& + \mathcal{D}_- \mathcal{M}_{+2}[2,2,\sqcup] \}, \\
		& \mathbb{M}_{43}[\sqcup] = - \mathcal{D}_+^{m_4 - n_3 -2} ( \eta_- \mathcal{M}_{+2}[2,1,\sqcup] - \eta_+ \mathcal{M}_{+2}[1,2,\sqcup] ), 
	\end{align}
	\begin{align}
		& \mathbb{M}_{14}[\sqcup] = \mathcal{D}_+^{m_1-n_4} ( P^2\mathcal{D}_- \mathcal{K}_- - \mathcal{P}_-^2 ) \mathcal{M}_{+2}[1,1,\sqcup], \\
		& \mathbb{M}_{24}[\sqcup] = \mathcal{D}_+^{m_2-n_4-1} \{ -\mathcal{K}_- ( \mathcal{M}_{+2}[2,1,\sqcup] + \mathcal{M}_{+2}[1,2,\sqcup] ) \nonumber\\
		& + \mathcal{P}_- ( \eta_-\mathcal{M}_{+2}[2,1,\sqcup] - \eta_+\mathcal{M}_{+2}[1,2,\sqcup] ) \}, \\
		& \mathbb{M}_{34}(\sqcup) = \mathcal{D}_+^{m_3-n_4-1} \{ \mathcal{P}_- ( \mathcal{M}_{+2}[2,1,\sqcup] + \mathcal{M}_{+2}[1,2,\sqcup] ) \nonumber\\
		& - P^2\mathcal{D}_- ( \eta_- \mathcal{M}_{+2}[2,1,\sqcup] - \eta_+ \mathcal{M}_{+2}[1,2,\sqcup] ) \}, \\
		& \mathbb{M}_{44}[\sqcup] = \mathcal{D}_+^{m_4 - n_4-1} \{ \{ \mathcal{K}_- + (\eta_+-\eta_-)\mathcal{P}_- - \eta_+ \eta_- P^2 \nonumber\\
		& \times \mathcal{D}_- \} \mathcal{M}_{+2}[1,1,\sqcup] + \mathcal{D}_- \mathcal{M}_{+2}[2,2,\sqcup] \}.
	\end{align}
	\end{subequations}
	Equation~\eqref{eq:def_mbb_M} contains the explicit form of linear operators to calculate the spectral functions of the BSWF. These identities also indicate the threshold $\Gamma_{\mathrm{th}}(z)$ in Eq.~\eqref{eq:NIR_BSWF_ps}.
	\section{Bethe--Salpeter equation\label{sc:BSE}}
	\subsection{Scalar decomposition}
	\begin{figure}
		\centering
		\includegraphics[width=\linewidth]{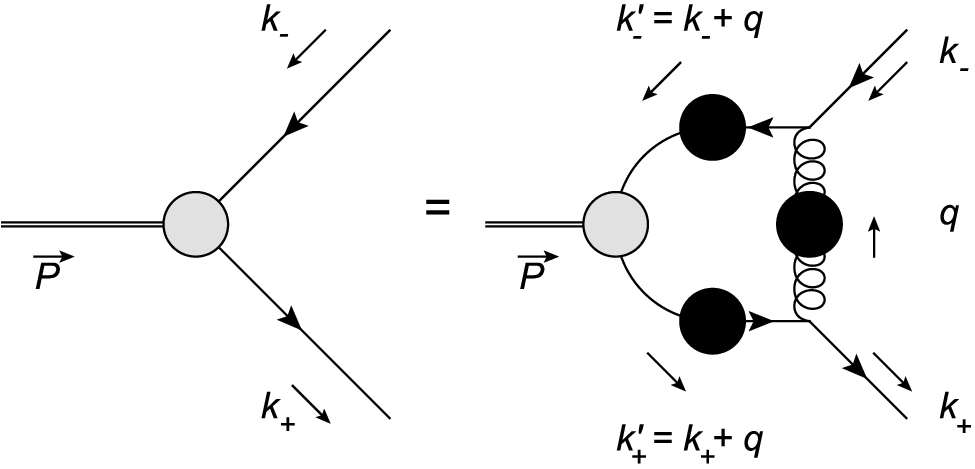}
		\caption{The BSE for the pseudoscalar bound state of a fermion and an antifermion in the rainbow-ladder truncation. The gray blobs and the black blobs are respectively the BSAs and the dressed propagators.}
		\label{fig:psbsefermions}
	\end{figure}
	Applying the rainbow-ladder truncation, the BSE for the amplitude of fermion-antifermion bound states is given by 
	\begin{align}
		& \Gamma(k,P) = -ig^2 \int d\underline{q} \, \gamma^\nu \chi(k+q,P) \gamma^\mu \nonumber\\
		& \times \left[ ( g_{\mu\nu} - q_\mu q_\nu / q^2 ) \, \mathcal{G}(q^2) + \xi q_\mu q_\nu / (q^2 + i\varepsilon)^2 \right], \label{eq:BSE_vertex}
	\end{align}
	where the integration measure is defined as
	\[ \int d\underline{q} = \dfrac{1}{(2\pi)^d} \int_{-\infty}^{+\infty}dq^0\prod_{j=1}^{d-1}\int_{-\infty}^{+\infty} dq^j \]
	with $d$ being the number of space-time dimensions that is set to $4$ after dimensional regularization~\cite{Jia:2024dfl}. This BSE is diagrammatically represented in Fig.~\ref{fig:psbsefermions} with $g$ being the coupling constant. Here $\mathcal{G}(q^2)$ is the dressing function of the gauge-boson propagator in the Landau gauge, the spectral representation of which is given by 
	\begin{equation}
		\mathcal{G}(q^2) = \int_{0}^{+\infty} dt \, \dfrac{ \rho_{\mathrm{g}}(t) }{ q^2-t+i\varepsilon },\label{eq:sr_gauge_boson_Landau}
	\end{equation}
	with $\rho_{\mathrm{g}}(t)$ being the spectral function. We adopt the covariant gauge for massless gauge bosons with $\xi$ being the gauge-fixing parameter. Because of the Ward--Green--Takahashi identity, the dressing of the gauge-boson propagator only applies to the Landau-gauge term~\cite{Peskin:1995ev}. This observation in conjunction with Eq.~\eqref{eq:sr_gauge_boson_Landau} allows the adaptation of the method developed in Ref.~\cite{Jia:2023imt} for massive exchange particles to the rainbow-ladder truncation with dressed massless gauge bosons. One may want to consider the renormalizable-$\xi$ gauge when the exchange gauge bosons are massive~\cite{Peskin:1995ev}.
	
	The spectral function in Eq.~\eqref{eq:sr_gauge_boson_Landau} remains unknown to the BSE, which is obtained either from the SDE for the guage-boson propagator or from a modeling of the interaction. After applying the Nakanishi integral representation of the BSWF in Eq.~\eqref{eq:NIR_BSWF_ps}, Eq.~\eqref{eq:BSE_vertex} becomes 
	\begin{align}
		& \Gamma(k,P) = -ig^2 \sum_{j=1}^{4} \int d\underline{q} \int_{z_{\mathrm{th}}^-}^{z_{\mathrm{th}}^+}dz \int_{\Gamma_{\mathrm{th}}(z)}^{+\infty} d\gamma \, \bigg\{  \nonumber\\
		& \int_{0}^{+\infty} dt \dfrac{ \gamma^\nu \gamma_5 T_j(k', P) \gamma^\mu ( g_{\mu\nu}- q_\mu q_\nu/q^2 )  \,\rho_{\mathrm{g}}(t) }{ [ k'^2 + zk'\cdot P - \gamma + i\varepsilon ]^{m_j} ( q^2 - t + i\varepsilon ) } \nonumber\\
		& + \dfrac{ \xi \,\slashed{q}\, \gamma_5 T_j(k', P)\, \slashed{q} }{ [ k'^2 + zk'\cdot P - \gamma + i\varepsilon ]^{m_j} ( q^2 + i\varepsilon )^2 } \bigg\} \, \Phi_j(\gamma,z) , \label{eq:BSE_ps_momentum}
	\end{align}
	where we have defined ${k' = k + q}$ and ${k'_\pm = k_\pm + q}$ for for notational conveniences. Due to the difference in the denominators, let us treat the contribution of the Landau-gauge exchange-boson propagator separately from that of the gauge-fixing term. 
	
	Explicitly the denominators associated with the Landau-gauge term can be combined using Feynman parametrization according to
	\begin{align}
		& \dfrac{1}{ [ k'^2 + zk'\cdot P - \gamma ]^{m_j} (q^2 - t + i\varepsilon ) } = m_j \int_0^1 dx \, x^{m_j-1} \nonumber\\
		& \times 1/ [ l^2 -\Delta(x,z,\gamma,t,k^2,k\cdot P) + i\varepsilon ]^{m_j+1}, \label{eq:FP_mj_qt}
	\end{align}
	with ${ l = q + x ( k + z P/2 ) }$ and
	\begin{align}
		& \Delta(x,z,\gamma,t,k^2, k\cdot P) = x\gamma + (1-x) t + x^2z^2 P^2/4 \nonumber\\
		& - x(1-x) (k^2 + zk\cdot P). \label{eq:def_common_dnm_Delta}
	\end{align}
	Let us define the following function
	\begin{equation}
		\mathcal{N}(\gamma_5 T_j(k',P)) = \gamma^\nu \gamma_5 T_j(k',P) \gamma^\mu ( g_{\mu\nu} - q_\mu q_\nu / q^2 )
	\end{equation}
	in order to decompose the result into scalar components. For each component we explicitly obtain
	\begin{subequations}\label{eq:momentum_bse_components_ps}
	\begin{align}
		& \mathcal{N}(\gamma_5) = \gamma_5(1-d), \\
		& \mathcal{N}(\gamma_5\slashed{P}) \rightarrow 
		\gamma_5 \bigg\{ \left[ d-3 + \dfrac{2l^2}{dq^2} + \dfrac{x^2z}{q^2} ( k\cdot P + zP^2/2 ) \right] \slashed{P} \nonumber\\
		& + \dfrac{x^2}{q^2} ( 2k\cdot P + z P^2 ) \slashed{k} \bigg\}, \\
		& \mathcal{N}(\gamma_5 \slashed{k}') \rightarrow \gamma_5 \bigg\{ xz\left[ \dfrac{1-d}{2} + \dfrac{x}{q^2} \left( k^2 + \dfrac{z}{2} k\cdot P \right) \right] \slashed{P} \nonumber\\
		& + \left[ d-3 + (1-d)x + \dfrac{ 2l^2 }{dq^2} + \dfrac{ x^2 }{q^2} ( 2k^2 + z k\cdot P ) \right] \slashed{k} \bigg\},
	\end{align}
	and 
	\begin{align}
		& \mathcal{N}(\gamma_5(-i)\sigma_{k'P}) \rightarrow \gamma_5 \big\{ x - x^2 ( k + z P / 2 )^2 / q^2 \nonumber\\
		& + (4-d) \left[ 1-x - l^2/(dq^2) \right] \big\} (-i)\sigma_{kP}. \label{eq:mathcal_N_sigma_kpp}
	\end{align}
	\end{subequations}
	Here the right arrows correspond to discarding terms odd in the integration momentum after shifts in the loop momentum. Having considered all terms of momenta in the numerator of the Landau-gauge contribution, in the last step of deriving Eq.~\eqref{eq:mathcal_N_sigma_kpp} we have applied
	\begin{equation}
		\int d\underline{l}\dfrac{1}{(l^2-\Delta)^{m_j}} \rightarrow \int d\underline{l} \dfrac{l^2 + x^2(k+zP/2)^2}{(l^2-\Delta)^{m_j} \, q^2}.\label{eq:loop_itg_dnm_q2}
	\end{equation}
	
	While unique to the contribution to the BSE from the $\xi$-dependent term of the gauge-boson propagator, another combination of denominators is given by
	\begin{align}
		& \dfrac{1}{(k'^2+zk'\cdot P - \gamma )^{m_j} ( q^2 +i\varepsilon )^2 } = m_j(m_j+1) \int_{0}^{1}dx \nonumber\\
		& \times x^{m_j-1}(1-x) / [l^2 - \Delta(x,z,\gamma,0,k^2,k\cdot P) + i\varepsilon ]^{m_j+2}, \label{eq:FP_mj_q2}
	\end{align}
	with the function $\Delta(x,z,\gamma,0,k^2,k\cdot P)$ as a special case of Eq.~\eqref{eq:def_common_dnm_Delta}. The shift in the loop momentum is identical to that for the Landau-gauge term. Similar to Eq.~\eqref{eq:momentum_bse_components_ps}, the momentum factors of the $\xi$-dependent term decompose into scalar components as follows:
	\begin{subequations}\label{eq:momentum_bse_gauge_components_ps}
	\begin{align}
		& \slashed{q} \gamma_5 \slashed{q} / q^2 = - \gamma_5, \\
		& \slashed{q} \gamma_5 \slashed{P} \slashed{q} / q^2 \rightarrow \gamma_5 \bigg\{ \left[ 1 - \dfrac{2l^2}{dq^2} - \dfrac{x^2z}{q^2}(k\cdot P + zP^2/2) \right] \slashed{P} \nonumber\\
		& - \dfrac{x^2}{q^2}(2k\cdot P + zP^2)\slashed{k} \bigg\}, \\
		& \slashed{q} \gamma_5 \slashed{k}' \slashed{q} / q^2 \rightarrow  \gamma_5 \bigg\{ xz\left[ \dfrac{1}{2} - \dfrac{x}{q^2} \left( k^2 + \dfrac{z}{2} k\cdot P \right) \right] \slashed{P} \nonumber\\
		& + \left[ x+1 - \dfrac{2l^2}{dq^2} - \dfrac{x^2}{q^2}(2k^2+zk\cdot P) \right] \slashed{k} \bigg\},
	\end{align}
	and
	\begin{align}
		& \slashed{q} \gamma_5 (-i)\sigma_{k'P}\, \slashed{q} / q^2 \rightarrow \gamma_5 \big[ -x + x^2(k+zP/2)^2 / q^2 \nonumber\\
		& + (4 - d) l^2 / (dq^2) \big] (-i)\sigma_{kP}, \label{eq:gauge_sigma_kpp}
	\end{align}
	\end{subequations}
	where we have applied Eq.~\eqref{eq:loop_itg_dnm_q2} in the final step of Eq.~\eqref{eq:gauge_sigma_kpp}.
	\subsection{Linear operators in scalar components of the BSE}
	In order to convert the BSE into integral equations for spectral functions, we would like to represent the momentum dependence on both sides of the identity using the Nakanishi integral representations similar to the result in Ref.~\cite{Jia:2023imt}. Specifically we encounter the following set of loop integrations on the right-hand side of Eq.~\eqref{eq:BSE_ps_momentum}:
	\begin{subequations}
	\begin{align}
		& -ig^2 \int d\underline{l}\, \dfrac{1}{ ( k'^2 + zk'\cdot P - \gamma )^m (q^2-t +i \varepsilon)}, \label{eq:BSE_loop_itg_1} \\
		& -ig^2 \int d\underline{l}\, \dfrac{1}{ ( k'^2 + zk'\cdot P - \gamma )^m (q^2-t +i \varepsilon)\,q^2}, \label{eq:BSE_loop_itg_2} \\
		& -ig^2 \int d\underline{l}\, \dfrac{ 2 l^2/d }{ ( k'^2 + zk'\cdot P - \gamma )^m (q^2-t +i \varepsilon) },  \label{eq:BSE_loop_itg_l2} \\
		& -ig^2\int d\underline{l}\, \dfrac{1}{(k'^2+zk'\cdot P - \gamma )^{m} ( q^2 +i\varepsilon )^2 }, \label{eq:BSE_loop_itg_q4} \\
		& -ig^2\int d\underline{l}\, \dfrac{2l^2/d}{(k'^2+zk'\cdot P - \gamma )^{m} ( q^2 +i\varepsilon )^2 }. \label{eq:BSE_loop_itg_l2_q4}
	\end{align}
	\end{subequations}
	Equations~\eqref{eq:BSE_loop_itg_1}, \eqref{eq:BSE_loop_itg_2}, and \eqref{eq:BSE_loop_itg_l2} are associated with the Landau-gauge boson propagator. Because Eq.~\eqref{eq:BSE_loop_itg_2} can be converted to Eq.~\eqref{eq:BSE_loop_itg_1} using 
	\begin{equation}
		\dfrac{1}{(q^2-t)\, q^2} = \dfrac{1}{t}\left( \dfrac{1}{q^2-t} - \dfrac{1}{q^2} \right),\label{eq:dnm_substitution}
	\end{equation}
	we only need to apply the Feynman parametrization in Eq.~\eqref{eq:FP_mj_qt} to combine momentum denominators in these terms. With $\Delta(x,z,\gamma,t,k^2,k\cdot P)$ given by Eq.~\eqref{eq:def_common_dnm_Delta}, let us then define 
	\begin{align}
		& \overline{\Delta}(x,z,\gamma,t,k^2,k\cdot P) = k^2 + {z k\cdot P} - \big\{ \gamma/(1-x) \nonumber\\
		& + t/x + xz^2P^2/[4(1-x)] \big\} \label{eq:def_overline_Delta}
	\end{align}
	such that ${\Delta(x,z,\gamma,t,k^2,k\cdot P) = -x(1-x) \overline{\Delta}}$. After applying Eq.~\eqref{eq:loop_int_dl}, Eq.~\eqref{eq:BSE_loop_itg_1} becomes 
	\begin{align}
		& -i \int d\underline{l} \dfrac{1}{ \left[ k'^2 + zk'\cdot P - \gamma \right]^m (q^2-t +i \varepsilon)} = -i \int d\underline{l} \nonumber\\
		& \int_{0}^{1}dx \dfrac{m\, x^{m-1}}{(l^2-\Delta)^{m+1}} = \dfrac{(-1)^{m+1}}{(4\pi)^2 (m-1)} \int_0^1 dx \, ( x/\Delta )^{m-1} \nonumber\\
		& = \dfrac{1}{(4\pi)^2(m-1)}\int_{0}^{1}dx \dfrac{1}{ [(1-x)\overline{\Delta}(x,z,\gamma,t,k^2,k\cdot P)]^{m-1} },\label{eq:BSE_loop_itg_1_eval}
	\end{align}
	where the function $\overline{\Delta}(x,z,\gamma,t,k^2,k\cdot P)$ captures the momentum dependence in the denominator after the loop integration. We would like to find a function $\varphi(\gamma,z,t)$ that satisfies
	\begin{align}
		& \int dz \int d\gamma\, \dfrac{\varphi(\gamma,z,t)}{(k^2+zk\cdot P - \gamma)^{m-1}} = \dfrac{1}{m-1} \int_{z_{\mathrm{th}}^-}^{z_{\mathrm{th}}^+}dz \nonumber\\
		& \int_{\Gamma_{\mathrm{th}}(z)}^{+\infty} d\gamma \int_0^1 dx\, \dfrac{\Phi(\gamma,z)}{[ (1-x) \overline{\Delta}(x,z,\gamma,t,k^2,k\cdot P) ]^{m-1} }, \label{eq:cond_functional_B1}
	\end{align}
	where integration limits on the left-hand side are implied from identities to be developed. Consequently $\varphi(\gamma,z,t)$ is related to $\Phi(\gamma,z)$ by 
	\begin{align}
		& \varphi(\gamma,z,t) = \dfrac{1}{m-1} \int_{z_{\mathrm{th}}^-}^{z_{\mathrm{th}}^+}dz' \int_{\Gamma_{\mathrm{th}}(z)}^{+\infty} d\gamma' \int_{0}^{1} dx \dfrac{ \delta(z-z') }{ (1-x)^{m-1} } \nonumber\\
		& \times \delta\left( \gamma - \left[ \dfrac{\gamma'}{1-x} + \dfrac{t}{x} + \dfrac{xz'^2P^2}{4(1-x)} \right] \right) \Phi(\gamma',z').
	\end{align}
	Here the $\delta$-function of $z-z'$ is trivial, indicating $z^{\pm}_{\mathrm{th}}$ as the limits of angular integration for both $\phi(\gamma,z,t)$ and $\Phi(\gamma,z)$. While applying procedures similar to those in Ref.~\cite{Jia:2023imt}, the $\delta$-function involving $\gamma$ can be reduced according to 
	\begin{align}
		& \delta\left(\gamma - \left[ \dfrac{\gamma'}{1-x} + \dfrac{t}{x} + \dfrac{xz'^2P^2}{4(1-x)} \right]\right) = x(1-x) \nonumber\\
		& \times \dfrac{ \delta(x-x_-(\gamma',z',\gamma,t)) + \delta(x-x_+(\gamma',z',\gamma,t)) } { \sqrt{ (\gamma + t -\gamma')^2 - (4\gamma + z'^2 P^2)t } },
	\end{align}
	with 
	\begin{align}
		& x_\pm(\gamma',z',\gamma,t) = \dfrac{1}{2\left(\gamma + z'^2 P^2/4 \right)} \big[ \gamma + t -\gamma' \nonumber\\
		& \pm \sqrt{ (\gamma+t-\gamma')^2 - (4\gamma+z'^2 P^2)t } \, \big]. \label{eq:def_x_pm}
	\end{align}
	Since at both ends of $x=0$ and $x=1$ the argument of this $\delta$-function as a quadratic function of $x$ is positive, the roots $x_\pm(\gamma',z',\gamma,t)$ are real and fall into $[0,1]$ only if 
	\begin{equation}
		t-\gamma -z'^2 P^2/2 \leq \gamma' \leq \gamma + t - \sqrt{ (4\gamma + z'^2 P^2 )t }. \label{eq:bound_gamma_prime}
	\end{equation}
	The left inequality is a consequence of $0\leq (x_+ + x_-)/2\leq 1$. While the right inequality comes from $(\gamma + t - \gamma')^2 - (4\gamma + z'^2 P^2)\,t \geq 0$. Equation~\eqref{eq:bound_gamma_prime} further suggests that 
	\[ \gamma + z'^2P^2/4 \geq t. \]
	When these conditions are satisfied, we obtain 
	\begin{subequations}\label{eq:def_functional_B1}
	\begin{align}
		& \varphi(\gamma,z,t) = \dfrac{1}{m-1}\int_{t-\gamma - z^2 P^2/2}^{\gamma + t - \sqrt{ (4\gamma + z^2 P^2)t }}d\gamma' \nonumber\\
		& \times \bigg\{ \dfrac{x_+(\gamma',z,\gamma,t)}{[1-x_+(\gamma',z,\gamma,t)]^{m-2}} + \dfrac{x_-(\gamma',z,\gamma,t)}{[1-x_-(\gamma',z,\gamma,t)]^{m-2}} \bigg\} \nonumber\\
		& \times \dfrac{ \theta( \gamma' - \Gamma_{\mathrm{th}}(z)) \, \Phi(\gamma',z)}{ \sqrt{ (\gamma+t-\gamma')^2 - (4\gamma + z^2 P^2)t } }, \label{eq:def_B1_itg_form}
	\end{align}
	similar to Eq.~(26) of Ref.~\cite{Jia:2023imt}. Equation~\eqref{eq:def_B1_itg_form} defines $\varphi(\gamma,z,t)$ as a linear functional $\mathcal{B}_{-}$ of $\Phi(\gamma,z)$ by
	\begin{equation}
		\varphi(\gamma,z,t) = \mathcal{B}_{-}[\Phi](\gamma,z,t).
	\end{equation}
	\end{subequations}
	Here the square-root singularity located at the upper integration limit of $\gamma'$ is integrable, the numerical evaluation of which can be carried applying a substitution of integration variable~\cite{Jia:2023imt}.
	
	Meanwhile the application of Eq.~\eqref{eq:loop_int_dl_l2} converts the loop integral in Eq~\eqref{eq:BSE_loop_itg_l2} according to
	\begin{align}
		& - i \int d\underline{l} \, \dfrac{ 2l^2/d}{ [k'^2 + zk'\cdot P -\gamma]^{m}(q^2-t+i\varepsilon) } \nonumber\\
		& = -\dfrac{2i}{d} \int d\underline{l} \int_{0}^{1}dx \dfrac{m\, x^{m-1} l^2}{(l^2-\Delta)^{m+1}} = \dfrac{(-1)^m}{(4\pi)^2(m-1)(m-2)} \nonumber\\
		& \times \int_{0}^{1}dx \dfrac{x^{m-1}}{\Delta^{m-2}} = \dfrac{1}{(4\pi)^2(m-1)(m-2)} \int_0^1 dx \nonumber\\
		& \times x/ [(1-x) \overline{\Delta}(x,z,\gamma,t,k^2,k\cdot P)]^{m-2}.
	\end{align}
	In this case we would like to find another function $\varphi(\gamma,z,t)$ that is related to $\Phi(\gamma,z)$ by 
	\begin{align}
		& \int dz \int d\gamma \dfrac{\varphi(\gamma,z,t)}{ (k^2 + z k\cdot P -\gamma )^{m-2}} = \dfrac{1}{(m-1)(m-2)} \nonumber\\
		& \times \int_{z_{\mathrm{th}}^-}^{z_{\mathrm{th}}^+}dz \int_{\Gamma_{\mathrm{th}}(z)}^{+\infty} d\gamma \int_{0}^{1}dx\, {x\,\Phi(\gamma,z)} \nonumber\\
		& \times 1/{ [ (1-x) \overline{\Delta}(x,z,\gamma,t,k^2,k\cdot P)]^{m-2} }. \label{eq:cond_functional_B2}
	\end{align}
	Again integration limits on the left-hand side are implied. Similar to the process in deriving Eq.~\eqref{eq:def_B1_itg_form}, we obtain
	\begin{subequations}\label{eq:def_functional_B2}
	\begin{align}
		& \varphi(\gamma,z,t) = \dfrac{ 1 }{ (m-1)(m-2) } \int_{t-\gamma-z^2P^2/2}^{\gamma+t-\sqrt{(4\gamma + z^2 P^2)t}} d\gamma' \nonumber\\
		& \times \bigg\{ \dfrac{x_+^2(\gamma',z,\gamma,t)}{[1-x_+(\gamma',z,\gamma,t)]^{m-3}} + \dfrac{x_-^2(\gamma',z,\gamma,t)}{[1-x_-(\gamma',z,\gamma,t)]^{m-3}} \Bigg\} \nonumber\\
		& \times \dfrac{ \theta( \gamma' - \Gamma_{\mathrm{th}}(z)) \, \Phi(\gamma',z)}{ \sqrt{ (\gamma+t-\gamma')^2 - (4\gamma + z^2 P^2)t } }. \label{eq:def_B2_itg_form}
	\end{align}
	Equation~\eqref{eq:def_B2_itg_form} defines a function $\varphi(\gamma,z,t)$ given by a linear functional $\mathcal{B}_{-2}$ of $\Phi(\gamma,z)$ according to
	\begin{equation}
	\varphi(\gamma,z,t) = \mathcal{B}_{-2}[\Phi ](\gamma,z,t)
	\end{equation}
	\end{subequations}
	such that Eq.~\eqref{eq:cond_functional_B2} is satisfied in the momentum space.
	
	As a generalization of Eqs.~\eqref{eq:def_functional_B1} and \eqref{eq:def_functional_B2}, let us define the functional $\mathcal{B}_{-\kappa}^{\nu}[\Phi](\gamma,z,t)$ with indices $\kappa$ and $\nu$ as
	\begin{align}
		& \varphi(\gamma,z,t) = \mathcal{B}_{-\kappa}^{\nu}[\Phi](\gamma,z,t) = \Gamma(m-\kappa) / \Gamma(m) \nonumber\\
		& \times \int_{t-\gamma - z^2 P^2/2}^{\gamma + t - \sqrt{(4\gamma + z^2 P^2)t}}d\gamma' \bigg\{ \dfrac{x_+^{\nu+\kappa}(\gamma',z,\gamma,t)}{[1-x_+(\gamma',z,\gamma,t)]^{m-\kappa -1}} \nonumber\\
		& + \dfrac{x_-^{\nu+\kappa}(\gamma',z,\gamma,t)}{[1-x_-(\gamma',z,\gamma,t)]^{m-\kappa -1}} \bigg\} \nonumber\\
		& \times \dfrac{ \theta( \gamma' - \Gamma_{\mathrm{th}}(z)) \, \Phi(\gamma',z)}{ \sqrt{ (\gamma+t-\gamma')^2 - (4\gamma + z^2 P^2)t } }, \label{eq:def_functional_B_kappa_nu}
	\end{align}
	where the associated Nakanishi integral representation applying $\varphi(\gamma,z,t)$ is given by 
	\begin{equation}
		\int_{z^-_{\mathrm{th}}}^{z^+_{\mathrm{th}}} dz \int d\gamma \dfrac{\varphi(\gamma,z,t)}{(k^2 + zk\cdot P - \gamma)^{m-\kappa}}. \label{eq:NIR_varphi}
	\end{equation}
	As discussed previously, the angular integration limits in Eq.~\eqref{eq:NIR_varphi} are identical to those of $\Phi(\gamma,z)$. While the radial integration limits in Eq.~\eqref{eq:NIR_varphi} are indicated by Eq.~\eqref{eq:bound_gamma_prime}. Equation~\eqref{eq:def_functional_B_kappa_nu} could be utilized to accommodate numerators with powers of $x$ other than instances included in Eqs.~\eqref{eq:cond_functional_B1} and \eqref{eq:cond_functional_B2}. Functionals $\mathcal{B}_{-}$ in Eq.~\eqref{eq:def_functional_B1} and $\mathcal{B}_{-2}$ in Eq.~\eqref{eq:def_functional_B2} are also respectively reproduced by $\mathcal{B}_{-1}^{0}$ and $\mathcal{B}_{-2}^{0}$ through Eq.~\eqref{eq:def_functional_B_kappa_nu}. 
	
	In order to calculate the contribution to the BSE from the $\xi$-dependent term of the gauge-boson propagator, after the loop integration we encounter momentum-space denominators identical to that given by Eq.~\eqref{eq:def_overline_Delta} with $t=0$. In this case the integration limits specified by Eq.~\eqref{eq:bound_gamma_prime} is reduced to
	\begin{equation}
		-\gamma - z'^2P^2/2 \leq \gamma' \leq \gamma.\label{eq:bound_gamma_prime_xi}
	\end{equation}
	Meanwhile the roots $x_{\pm}$ defined in Eq.~\eqref{eq:def_x_pm} become
	\begin{subequations}
	\begin{align}
		& x_+(\gamma',z',\gamma,0) = (\gamma - \gamma')/(\gamma + z'^2 P^2/4), \\
		& x_-(\gamma',z',\gamma,0) = 0.
	\end{align}
	\end{subequations}
	This limit is also reached when Eq.~\eqref{eq:dnm_substitution} is applied to convert the loop integral in Eq.~\eqref{eq:BSE_loop_itg_2} into that in Eq.~\eqref{eq:BSE_loop_itg_1} when calculating the contribution from the Landau-gauge boson propagator. Within this limit we have the reduced case of $\mathcal{B}^{\nu}_{-\kappa}[\sqcup](\gamma,z,t)$ defined in Eq.~\eqref{eq:def_functional_B_kappa_nu} that gives a spectral function $\phi(\gamma,z)$ from a linear functional of $\Phi(\gamma,z)$ formally as
	\begin{subequations}\label{eq:def_functional_B_kappa_nu_t0}
	\begin{equation}
		\phi(\gamma,z) = \mathcal{B}^{\nu}_{-\kappa}[\Phi](\gamma,z,0). 
	\end{equation}
	The functional $\mathcal{B}^{\nu}_{-\kappa}[\sqcup](\gamma,z,0)$ is only well-defined when ${ \nu+\kappa > 0 }$ to avoid singularities due to the integration near $\gamma'=\gamma$. When such a condition is met, Eq.~\eqref{eq:def_functional_B_kappa_nu} is reduced into
	\begin{align}
		& \phi(\gamma,z) = \dfrac{\Gamma(m-\kappa)}{\Gamma(m)}\int_{-\gamma - z^2 P^2/2}^{\gamma}d\gamma' \, (\gamma-\gamma')^{\nu+\kappa-1} \nonumber\\
		& \times \dfrac{ (\gamma+z^2P^2/4)^{m-2\kappa-\nu-1} }{ (\gamma'+z^2P^2/4)^{m-\kappa-1} } \theta(\gamma'-\Gamma_{\mathrm{th}}(z))\, \Phi(\gamma',z).
	\end{align}
	\end{subequations}
	
	Specifically for the calculation of the contribution from the $\xi$-dependent term, based on Eq.~\eqref{eq:BSE_loop_itg_1_eval} we have
	\begin{align}
		& -i\int d\underline{l}\dfrac{1}{(k'^2+zk'\cdot P - \gamma )^{m} ( q^2 +i\varepsilon ) } = \dfrac{1}{(4\pi)^2(m-1)} \nonumber\\
		& \times \int_{0}^{1}dx \dfrac{1}{[ (1-x)\overline{\Delta}(x,z,\gamma,0,k^2,k\cdot P)]^{m-1}},
	\end{align}
	which corresponds to finding a function $\phi(\gamma,z)$ that satisfies
	\begin{align}
		& \int dz \int d\gamma\, \dfrac{\phi(\gamma,z)}{(k^2+zk\cdot P - \gamma)^{m-1}} = \dfrac{1}{m-1} \int_{z_{\mathrm{th}}^-}^{z_{\mathrm{th}}^+}dz \nonumber\\
		& \times \int_{\Gamma_{\mathrm{th}}(z)}^{+\infty} d\gamma \int_0^1 dx \dfrac{\Phi(\gamma,z)}{[ (1-x) \overline{\Delta}(x,z,\gamma,0,k^2,k\cdot P) ]^{m-1} } \label{eq:cond_functional_B1_q2}
	\end{align}
	in the momentum space. Such a function is given by 
	\begin{equation}
		\varphi(\gamma,z) = \mathcal{B}^{\nu}_{-1}[\Phi](\gamma,z,0) \label{eq:form_functional_B_1_nu_t0}
	\end{equation}
	with $\nu=0$. Due to the occurrence of the Feynman parameter $x$ in the numerators we also need other cases of $\mathcal{B}^{\nu}_{-1}[\sqcup](\gamma,z,0)$ for ${\nu > -1}$. The explicit form of Eq.~\eqref{eq:form_functional_B_1_nu_t0} is given by a special case of Eq.~\eqref{eq:def_functional_B_kappa_nu_t0}.
	
	In the meantime we need to calculate the integrals in Eqs.~\eqref{eq:BSE_loop_itg_q4} and \eqref{eq:BSE_loop_itg_l2_q4}. Specifically the combination of Eqs.~\eqref{eq:FP_mj_q2} and \eqref{eq:loop_int_dl_q2} gives
	\begin{align}
		& -i\int d\underline{l}\dfrac{1}{(k'^2+zk'\cdot P - \gamma )^{m} ( q^2 +i\varepsilon )^2 } = -i\int d\underline{l}\int_{0}^{1}dx \nonumber\\
		& \times \dfrac{m(m+1)x^{m-1}(1-x)}{ [l^2 - \Delta(x,z,\gamma,0,k^2,k\cdot P) ]^{m+2} } = \dfrac{ (-1)^m }{(4\pi)^2} \int_{0}^{1}dx\, x^{m-1} \nonumber\\
		& \times \dfrac{ 1-x }{ [ \Delta(x,z,\gamma,0,k^2,k\cdot P) ]^{m} } = \dfrac{1}{(4\pi)^2} \int_{0}^{1}dx \dfrac{1}{ x (1-x)^{m-1} } \nonumber\\
		& \times [ \overline{\Delta}(x,z,\gamma,0,k^2,k\cdot P) ]^{-m} ,\label{eq:dnm_functional_A_q2x2}
	\end{align}
	which corresponds to finding a function $\phi(\gamma,z)$ such that
	\begin{align}
		& \int dz \int d\gamma \dfrac{\phi(\gamma,z)}{(k^2 + zk\cdot P -\gamma)^m} = \int_{ z^-_{\mathrm{th}} }^{z^+_{\mathrm{th}}}dz \int_{\Gamma_{\mathrm{th}}(z)}^{+\infty}d\gamma \nonumber\\
		& \times \dfrac{ x\, \Phi(\gamma,z) }{ (1-x)^{m-1} } [ \overline{\Delta}(x,z,\gamma,0,k^2,k\cdot P) ]^{-m}.\label{eq:ori_functional_A_q2x2}
	\end{align}
	Here we have multiplied a factor of $x^2$ to the numerator on the right-hand side of Eq.~\eqref{eq:dnm_functional_A_q2x2} due to its association with relevant terms in Eq.~\eqref{eq:momentum_bse_gauge_components_ps}. Explicitly when the function $\phi(\gamma,z)$ is given by
	\begin{subequations}\label{eq:def_functional_A_q2x2}
	\begin{align}
		& \phi(\gamma,z) = \int_{-\gamma-z^2P^2/2}^{\gamma} d\gamma'\, ( \gamma-\gamma' ) \dfrac{(\gamma+z^2P^2/4)^{m-4}}{(\gamma'+z^2P^2/4)^{m-2}} \nonumber\\
		& \times \theta(\gamma'-\Gamma_{\mathrm{th}}(z))\,\Phi(\gamma,z), \label{eq:def_A_itg_form_q4x2}
	\end{align}
	Eq.~\eqref{eq:ori_functional_A_q2x2} is satisfied in the momentum space. Although Eq.~\eqref{eq:def_A_itg_form_q4x2} is very close to the special case of Eq.~\eqref{eq:def_functional_B_kappa_nu_t0} with ${\nu = \kappa = 1}$, it does not change the indices of denominators in the Nakanishi integral representations. Therefore Eq.~\eqref{eq:def_A_itg_form_q4x2} defines anther functional $\mathcal{A}_0$ formally as
	\begin{equation}
		\phi(\gamma,z) = \mathcal{A}_0[\Phi](\gamma,z).
	\end{equation}
	\end{subequations}
	Additionally based on Eqs.~\eqref{eq:FP_mj_q2} and \eqref{eq:loop_int_dl_l2_q2} we have
	\begin{align}
		& -i\int d\underline{l}\dfrac{2l^2/d}{(k'^2+zk'\cdot P - \gamma )^{m} ( q^2 +i\varepsilon )^2 } = - \dfrac{2i}{d} \int d\underline{l} \nonumber\\
		& \times \int_{0}^{1}dx \dfrac{m(m+1)x^{m-1}(1-x)\, l^2 }{ [l^2 - \Delta(x,z,\gamma,0,k^2,k\cdot P) ]^{m+2} } = \dfrac{1}{(4\pi)^2} \int_{0}^{1}dx \nonumber\\
		& \times \dfrac{(-1)^{m+1} x^{m-1}(1-x) }{(m-1)[\Delta(x,z,\gamma,0,k^2,k\cdot P)]^{m-1}} = \dfrac{1}{(4\pi)^2 (m-1) } \nonumber\\
		& \times \int_{0}^{1}dx \dfrac{ 1 }{ (1-x)^{m-2} [\overline{\Delta}(x,z,\gamma,0,k^2,k\cdot P)]^{m-1} } 
	\end{align}
	which corresponds to finding a function $\phi(\gamma,z)$ such that
	\begin{align}
		& \int dz \int d\gamma \dfrac{\phi(\gamma,z)}{(k^2 + zk\cdot P -\gamma)^{m-1}} = \dfrac{1}{m-1}\int_{ z^-_{\mathrm{th}} }^{z^+_{\mathrm{th}}}dz \nonumber\\
		& \times \int_{\Gamma_{\mathrm{th}}(z)}^{+\infty}d\gamma \dfrac{ \Phi(\gamma,z) }{ (1-x)^{m-2} [ \overline{\Delta}(x,z,\gamma,0,k^2,k\cdot P) ]^{m-1} }
	\end{align}
	is satisfied in the momentum space. Such a function is given by
	\begin{subequations}\label{eq:def_functional_An1}
	\begin{align}
		& \phi(\gamma,z) = \dfrac{1}{m-1} \int_{-\gamma-z^2P^2/2}^{\gamma} d\gamma' \, \dfrac{ (\gamma+z^2P^2/4)^{m-4} }{ (\gamma'+z^2P^2/4)^{m-3} } \nonumber\\
		& \times \theta(\gamma'-\Gamma_{\mathrm{th}}(z))\,\Phi(\gamma,z),\label{eq:def_An1_itg_form}
	\end{align}
	which is again reproduced by Eq.~\eqref{eq:def_functional_B_kappa_nu_t0} when ${\nu = -1}$ and ${\kappa = 2}$. Because the index of the denominator in the Nakanishi integral representation using $\Phi(\gamma,z)$ is reduced by $1$ instead of $2$, Eq.~\eqref{eq:def_An1_itg_form} however defines another functional $\mathcal{A}_{-1}$ formally as
	\begin{equation}
		\phi(\gamma,z) = \mathcal{A}_{-1}[\Phi](\gamma,z).
	\end{equation}
	\end{subequations}
	We have derived all linear operators of the Nakanishi spectral functions needed to mach each of the components in the scalar decomposition of the BSE.
	\subsection{BSE as a matrix of linear functionals}
	Having derived the linear operators that match the momentum-space behavior for all scalar components of the BSE in the previous subsection, the BSE in the form of Eq.~\eqref{eq:BSE_ps_momentum} indicates the following relation:
	\begin{subequations}\label{eq:def_mbb_B}
	\begin{equation}
		\phi_j(\gamma,z) = \dfrac{g^2}{(4\pi)^2} \sum_{k=1}^{4} \mathbb{B}_{jk}[\Phi_k](\gamma,z),
	\end{equation}
	or in matrix form as
	\begin{equation*}
		\begin{pmatrix}
			\phi_1(\gamma,z)\\
			\phi_2(\gamma,z)\\
			\phi_3(\gamma,z)\\
			\phi_4(\gamma,z)
		\end{pmatrix} 
		= \lambda
		\begin{pmatrix}
			\mathbb{B}_{11}[] & \mathbb{B}_{12}[] & \mathbb{B}_{13}[] & \mathbb{B}_{14}[] \\
			\mathbb{B}_{21}[] & \mathbb{B}_{22}[] & \mathbb{B}_{23}[] & \mathbb{B}_{24}[] \\
			\mathbb{B}_{31}[] & \mathbb{B}_{32}[] & \mathbb{B}_{33}[] & \mathbb{B}_{34}[] \\
			\mathbb{B}_{41}[] & \mathbb{B}_{42}[] & \mathbb{B}_{43}[] & \mathbb{B}_{44}[]
		\end{pmatrix}
		\begin{pmatrix}
			\Phi_1(\gamma,z) \\
			\Phi_2(\gamma,z) \\
			\Phi_3(\gamma,z) \\
			\Phi_4(\gamma,z)
		\end{pmatrix},
	\end{equation*}
	where $\mathbb{B}_{jk}$ is an array of linear operators on the spectral functions $\Phi_k(\gamma,z)$ of the BSWF and $\rho_{\mathrm{g}}(t)$ of the gauge-boson propagator. We have also introduced $\lambda= g^2/(4\pi)^2$ as the eigenvalue of the BSE. Specifically base on Eqs.~\eqref{eq:momentum_bse_components_ps} and \eqref{eq:momentum_bse_gauge_components_ps} we obtain 
	\begin{align}
		& \mathbb{B}_{11}[\sqcup] = -\mathcal{D}_+^{n_1-m_1+1} \Big( 3\int_{0}^{+\infty} dt \,\rho_{\mathrm{g}}(t) \mathcal{B}_{-1}^{0}[\sqcup](\bullet,\bullet,t) \nonumber\\
		& + \xi \, \mathcal{B}_{-1}^{0}[\sqcup](\bullet,\bullet,0) \Big),
	\end{align}
	$\mathbb{B}_{21}[\sqcup] = \mathbb{0}$, $\mathbb{B}_{31}[\sqcup] = \mathbb{0}$, and $\mathbb{B}_{41}[\sqcup] = \mathbb{0}$. Here we have defined the rule for filling functions and variables as ${\mathcal{B}[\sqcup](\bullet,\bullet,t)\Phi(\gamma,z) = \mathcal{B}[\Phi](\gamma,z,t)}$ in addition to Eq.~\eqref{eq:def_spaceholders}. While for other components we obtain
	\begin{align}
		& \mathbb{B}_{22}[\sqcup] = \mathcal{D}_+^{n_2-m_2+1} \bigg\{ \int_{0}^{+\infty} dt\, \rho_{\mathrm{g}}(t) \Big\{ \mathcal{B}_{-1}^{0}[\sqcup](\bullet,\bullet,t) \nonumber\\
		& + \dfrac{1}{t} \Big( \mathcal{D}_+ \big( \mathcal{B}_{-2}^{0}[\sqcup](\bullet,\bullet,t) + \mathcal{P}_- \mathcal{B}_{-1}^{2}[z\sqcup](\bullet,\bullet,t) \big) + \dfrac{P^2}{2} \nonumber\\
		& \times \mathcal{B}_{-1}^{2}[z^2\sqcup](\bullet,\bullet,t) - \left(t\rightarrow 0 \right) \Big) \Big\} + \xi \Big( \mathcal{B}^{0}_{-1}[\sqcup](\bullet,\bullet,0) \nonumber\\
		& - \mathcal{A}_{-1}[\sqcup] - \mathcal{P}_{-}\mathcal{A}_{0}[z\sqcup] - \dfrac{P^2}{2} \mathcal{D}_{-} \mathcal{A}_{0}[z^2\sqcup] \Big) \bigg\}, \\
		& \mathbb{B}_{32}[\sqcup] = \mathcal{D}_+^{n_3-m_2+1} \Big\{ \int_{0}^{+\infty} dt\, \dfrac{\rho_{\mathrm{g}}(t)}{t} \big\{ 2\mathcal{D}_+ \mathcal{P}_- \nonumber\\
		& \times \mathcal{B}_{-1}^{2}[\sqcup](\bullet,\bullet,t) + P^2 \mathcal{B}_{-1}^{2}[z\sqcup](\bullet,\bullet,t) - (t\rightarrow 0) \big\} \nonumber\\
		& - \xi\big( 2\mathcal{P}_{-} \mathcal{A}_{0}[\sqcup] + P^2 \mathcal{D}_{-} \mathcal{A}_{0}[z\sqcup] \big) \Big\},
	\end{align}
	$\mathbb{B}_{12}[\sqcup] = \mathbb{0}$, and $\mathbb{B}_{42}[\sqcup] = \mathbb{0}$. We also obtain
	\begin{align}
		& \mathbb{B}_{23}[\sqcup] = \mathcal{D}_+^{n_2-m_3+1} \bigg\{ \int_0^{+\infty} dt\, \rho_{\mathrm{g}}(t) \Big\{ -\dfrac{3}{2} \mathcal{B}_{-1}^{1}[z\sqcup](\bullet,\bullet,t) \nonumber\\
		& + \dfrac{\mathcal{D}_+}{t} \Big( \mathcal{K}_- \mathcal{B}_{-1}^{2}[z\sqcup](\bullet,\bullet,t) + \dfrac{1}{2} \mathcal{P}_- \mathcal{B}_{-1}^{2}[z^2\sqcup](\bullet,\bullet,t) \nonumber\\
		& - (t\rightarrow 0) \Big) \Big\} + \xi \Big( \dfrac{1}{2}\mathcal{B}_{-1}^{1}[z\sqcup](\bullet,\bullet,0) - \mathcal{K}_{-}\mathcal{A}_{0}[z\sqcup] \nonumber\\
		& - \dfrac{1}{2}\mathcal{P}_{-} \mathcal{A}_{0}[z^2\sqcup] \Big) \bigg\}, \\
		& \mathbb{B}_{33}[\sqcup] = \mathcal{D}_+^{n_3-m_3+1} \bigg\{ \int_0^{+\infty} dt\, \rho_{\mathrm{g}}(t) \Big\{ \mathcal{B}_{-1}^{0}[\sqcup](\bullet,\bullet,t) \nonumber\\
		& - 3 \mathcal{B}_{-1}^{1}[\sqcup](\bullet,\bullet,t) + \dfrac{\mathcal{D}_+}{t} ( \mathcal{B}_{-2}^{0}[\sqcup](\bullet,\bullet,t) + 2 \mathcal{K}_- \nonumber\\
		& \times \mathcal{B}_{-1}^{2}[\sqcup](\bullet,\bullet,t) + \mathcal{P}_- \mathcal{B}_{-1}^{2}[z\sqcup](\bullet,\bullet,t) - (t\rightarrow 0) ) \Big\} \nonumber\\
		& + \xi \Big( \mathcal{B}_{-1}^{0}[\sqcup](\bullet,\bullet,0) + \mathcal{B}_{-1}^{1}[\sqcup](\bullet,\bullet,0) - \mathcal{A}_{-1}[\sqcup] \nonumber\\
		& - 2 \mathcal{K}_{-}\mathcal{A}_{0}[\sqcup] - \mathcal{P}_{-}\mathcal{A}_{0}[z\sqcup] \Big) \bigg\}.
	\end{align}
	with $\mathbb{B}_{43}[\sqcup] = \mathbb{0}$ and $\mathbb{B}_{13}[\sqcup] = \mathbb{0}$. While for the $4$-th component of the BSWF we have $\mathbb{B}_{14}[\sqcup] = \mathbb{0}$, $\mathbb{B}_{24}[\sqcup] = \mathbb{0}$, $\mathbb{B}_{34}[\sqcup] = \mathbb{0}$, and
	\begin{align}
		& \mathbb{B}_{44}[\sqcup] = \mathcal{D}_+^{n_4-m_4+1} \bigg\{ \int_0^{+\infty} dt\, \rho_{\mathrm{g}}(t) \Big\{ \mathcal{B}_{-1}^{1}[\sqcup](\bullet,\bullet,t) \nonumber\\
		& - \dfrac{1}{t} \Big( \mathcal{D}_+ \big( \mathcal{K}_-\mathcal{B}_{-1}^{2}[\sqcup](\bullet,\bullet,t) + \mathcal{P}_-\mathcal{B}_{-1}^{2}[z\sqcup](\bullet,\bullet,t) \big) + \dfrac{P^2}{4} \nonumber\\
		& \times \mathcal{B}_{-1}^{2}[z^2\sqcup](\bullet,\bullet,t) - (t\rightarrow 0) \Big) \Big\} + \xi \Big( - \mathcal{B}_{-1}^{1}[\sqcup](\bullet,\bullet,0) \nonumber\\
		&  + \mathcal{K}_{-}\mathcal{A}_{0}[\sqcup] + \mathcal{P}_{-}\mathcal{A}_{0}[z\sqcup] + \dfrac{P^2}{4}\mathcal{D}_{-}\mathcal{A}_{0}[z^2\sqcup] \Big) \bigg\}.
	\end{align}
	\end{subequations}
	\section{Summary and outlook\label{sc:summary}}
	Based on the tested method of solving for the Nakanishi spectral functions from the BSE in the massive variant of the Wick-Cutkosky model~\cite{Jia:2023imt}, we have developed the Minkowski-space formulation of the BSE for pseudoscalar bound states of a fermion and an antifermion in the rainbow-ladder truncation. Specifically after introducing the Nakanishi integral representations of the BSA and of the BSWF, we have first derived elementary operations on the spectral functions that correspond to multiplications of momentum factors. Based on the definition of the BSWF in the momentum space, we have then obtained Eq.~\eqref{eq:def_mbb_M} as the explicit expression for the Nakanishi spectral functions of the BSWF in terms of those of the BSA and the spectral functions of the fermion propagators. Meanwhile in the rainbow-ladder truncation in the covariant gauge for massless gauge bosons, we have derived Eq.~\eqref{eq:def_mbb_B} that gives the Nakanishi spectral functions of the BSA from its BSE with the assistance of the spectral representation for the exchange-boson propagator in the Landau gauge. In deriving these equations, we have also allowed the fermion propagators to be fully dressed. Both the kernel functions and the limits of integrations of these equations have been analytical functions. 
	
	After knowing the spectral functions of the propagators, Eqs.~\eqref{eq:def_mbb_M} and \eqref{eq:def_mbb_B} form a closed set of integral equations for Nakanishi spectral functions $\phi_j(\gamma,z)$ and $\Phi_j(\gamma,z)$ in Eqs.~\eqref{eq:NIR_BSA_ps} and \eqref{eq:NIR_BSWF_ps}. Specifically in terms of the the spectral functions of the BSA, we have converted the BSE into the following linear eigenvalue problem
	\begin{equation}
		\phi_i(\gamma,z) = \lambda \sum_{j=1}^{4}\sum_{k=1}^{4}\mathbb{B}_{ij}[\mathbb{M}_{jk}[\phi_k]](\gamma,z).
	\end{equation}
	A similar identity can also be obtained for the spectral functions of the BSWF. The current development therefore lays the foundation for an iterative solver of the BSE in the Minkowski space in conjunction with the SDEs for the propagators within the same truncation.
	We expect to apply the method developed here to solve the BSEs for QED bound states of leptons and for mesons within a modeling of QCD interactions. In order to minimize the number of the operation $\mathcal{D}_{+}$ in both Eqs.~\eqref{eq:def_mbb_M} and \eqref{eq:def_mbb_B}, we prefer to choose ${m_j = n_j + 1}$, ${n_2 = n_3}$, and ${n_4 = n_1+1}$ for the indices in the Nakanishi integral representations. Meanwhile based on the number of momentum dimensions for the Dirac bases in Eq.~\eqref{eq:def_T_j_basis}, we could set ${n_1 = 1}$ in Eq.~\eqref{eq:NIR_BSA_ps}. One would optionally consider letting ${n_2 = n_1}$ or ${n_2 = n_1 + 1}$ such that no free indices remain. 
	\begin{acknowledgments}
		We appreciate Dr. Ian Clo\"{e}t and Prof. Pieter Maris for discussions on this topic. This work was supported by the US Department of Energy, Office of Science, Office of Nuclear Physics, under Contract No. DE-AC02-06CH11357. 
	\end{acknowledgments}
	\appendix*
	\section*{Spectral representation of the fermion propagator}
	The fermion propagator can be decomposed into two Dirac components
	\begin{equation}
		S_{\mathrm{F}}(k) = \slashed{k}S_1(k^2) + S_2(k^2), \label{eq:dec_S_F}
	\end{equation}
	each of which yields the following K\"{a}ll\'{e}n-Lehmann spectral representation
	\begin{equation}
		S_j(k^2) = \int_{s_{\mathrm{th}}}^{+\infty}ds \dfrac{\rho_j(s)}{k^2-s+i\varepsilon} \label{eq:dec_S_Fj}
	\end{equation}
	for $j\in\{1,2\}$ with $s_{\mathrm{th}}$ being the threshold of the integration~\cite{Jia:2016wyu,Jia:2017niz}. The spectral functions in Eq.~\eqref{eq:dec_S_Fj} can be combined as
	\begin{equation}
		\rho(W) = \mathrm{sign}(W)[W\rho_1(W^2) + \rho_2(W^2) ],
	\end{equation}
	which is applied to give Eq.~\eqref{eq:SR_S_F}. When $W_{\mathrm{th}} = \sqrt{s_{\mathrm{th}}}$ Eq.~\eqref{eq:SR_S_F} is equivalent to Eq.~\eqref{eq:dec_S_F} given that the integration in $W$ can be written as
	\[ \int_{\vert W\vert \geq W_{\mathrm{th}}}^{+\infty} dW = \int_{-\infty}^{-W_{\mathrm{th}}}dW+ \int_{W_{\mathrm{th}}}^{+\infty}dW. \]
	After defining $s = W^2$ we subsequently obtain
	\begin{align}
		& S_{\mathrm{F}}(k) = \bigg\{ \int_{W_{\mathrm{th}}}^{+\infty}dW \dfrac{1}{\slashed{k}-W + i\varepsilon} - \int_{-\infty }^{-W_{\mathrm{th}}}dW \nonumber\\
		& \times \dfrac{1}{ \slashed{k}-W - i\varepsilon } \bigg\} [ W\rho_1(W^2) + \rho_2(W^2) ] = \int_{W_{\mathrm{th}}}^{+\infty}dW \nonumber\\
		& \times \left[ \dfrac{ W\rho_1(W^2) + \rho_2(W^2) }{\slashed{k}-W + i\varepsilon} + \dfrac{W\rho_1(W^2)-\rho_2(W^2)}{\slashed{k} + W - i\varepsilon} \right] \nonumber\\
		& = \int_{s_{\mathrm{th}}}^{+\infty}ds \, [ \slashed{k}\rho_1(s)+\rho_2(s) ] / (k^2-s+i\varepsilon)
	\end{align}
	that reproduces the spectral representation of the scalar component $S_j(k^2)$ in Eq.~\eqref{eq:dec_S_Fj}.
	\section*{Integration of the loop momentum}
	Applying dimensional regularization with $d=4-2\epsilon$ being the number of spacetime dimensions~\cite{Peskin:1995ev}, we have 
	\begin{align}
		& \int d\underline{l} \dfrac{l^{2m}}{(l^2-\Delta+i\varepsilon)^{n}} = \dfrac{i(-1)^{n+m}\Gamma(d/2+m)}{(4\pi)^{d/2}\Gamma(d/2)\Gamma(n)\Delta^{n-m-d/2}} \nonumber\\
		& \times \Gamma(n-m-d/2).
	\end{align}
	Ultraviolet divergence is not encountered in the BSE with our choice of $n_j\geq 1$ in Eq.~\eqref{eq:NIR_BSA_ps} when $d=4$. Specifically we obtain the following integrations in the BSE:
	\begin{subequations}
	\begin{align}
		& \int d\underline{l}\dfrac{1}{(l^2-\Delta)^{m+1}} = \dfrac{i(-1)^{m+1}}{(4\pi)^2 m (m-1) \Delta^{m-1}}, \label{eq:loop_int_dl} \\
		& \int d\underline{l}\dfrac{2l^2/d}{(l^2-\Delta)^{m+1}}	= \dfrac{i(-1)^m}{ (4\pi)^2 m(m-1)(m-2)\Delta^{m-2}}. \label{eq:loop_int_dl_l2}
	\end{align}
	Based on Eqs~\eqref{eq:loop_int_dl} and \eqref{eq:loop_int_dl_l2} we also have
	\begin{align}
		& \int d\underline{l}\dfrac{1}{(l^2-\Delta)^{m+2}} = \dfrac{i(-1)^m}{(4\pi)^2\, m \, (m+1)\, \Delta^{m}}, \label{eq:loop_int_dl_q2} \\
		& \int d\underline{l}\dfrac{2l^2/d}{(l^2-\Delta)^{m+2}} = \dfrac{i(-1)^{m+1}}{(4\pi)^2(m+1)m(m-1)\Delta^{m-1}}. \label{eq:loop_int_dl_l2_q2}
	\end{align}
	\end{subequations}
	\bibliography{bse_ps_refs}
\end{document}